%% file: main_arxiv.tex
\documentclass[runningheads]{llncs}
\input{macro.tex}
\begin{document}
\title{Oracle-Supported Dynamic Exploit Generation for Smart Contracts}
%
%
\author{Haijun Wang\inst{1} \and
Yi Li\inst{1} \and Shang-Wei Lin\inst{1}\and Cyrille Artho\inst{2}\and Lei Ma\inst{3} \and Yang Liu\inst{1}}
%
%
\institute{Nanyang Technological University, Singapore\\
\email{\{haijun.wang, yi\_li, shang-wei.lin, yangliu\}@ntu.edu.sg} \and
KTH Royal Institute of Technology, Sweden \\
\email{artho@kth.se} \and 
Kyushu University, Japan \\
\email{malei@ait.kyushu-u.ac.jp}
}
\maketitle              
\begin{abstract}
	Despite the high stakes involved in smart contracts, they are often developed in an 
	undisciplined manner, leaving the security and reliability of blockchain transactions at risk.
	In this paper, we introduce \tool\ --- an oracle-supported dynamic exploit generation framework 
	for smart contracts.
	Existing approaches mutate only single transactions;
	\tool exceeds these by mutating the transaction sequences.
	\tool uses data-flow, control-flow, and the dynamic contract state to guide its 
	mutations.
	It then monitors the executions of target contract programs, and validates the results against a
	general-purpose \SemanticTestOracle to discover vulnerabilities.
	Being a dynamic technique, it guarantees that each discovered vulnerability is a violation of the
	test oracle and is able to generate the attack script to exploit this vulnerability.
	In contrast to rule-based approaches, \tool has not shown any false positives, and it 
	easily generalizes to unknown types of vulnerabilities (e.g., logic errors).
	We evaluate \tool on 218 vulnerable smart contracts. The experimental results confirm its 
	practical applicability and advantages over the state-of-the-art techniques,
	and also reveal three new types of attacks.
\end{abstract}

\input{intro}
\input{background}
\input{oracle}
\input{fuzzing}
\input{eval}
\input{relate}
\input{conclusion}

%
%
%
%
\bibliographystyle{IEEEtran}
\bibliography{references}
\end{document}

%% file: macro.tex
\usepackage{xcolor}
\usepackage{graphicx}
\usepackage{booktabs}
\usepackage{microtype}
\usepackage{amsfonts}
\usepackage{multirow}
\usepackage{xspace}
\usepackage{amssymb}
\usepackage{wasysym}
\usepackage{color}
\usepackage{enumitem,etoolbox}
\usepackage{array}
\usepackage{microtype}
\usepackage{comment}
\usepackage[ruled,vlined]{algorithm2e}
\usepackage{xcolor}
\AtBeginEnvironment{verbatim}{\setlist[trivlist]{nosep}}
\usepackage[hyphens]{url}
\usepackage[hidelinks]{hyperref}
\hypersetup{breaklinks=true}
\newcommand{\tool}{ContraMaster\xspace}
\newcommand{\random}{ContraAFL\xspace}

\newcommand{\ether}{\texttt{Ether}\xspace}
\newcommand{\wei}{\texttt{Wei}\xspace}

\newcommand{\SemanticOracle}{semantic oracle\xspace}
\newcommand{\SemanticTestOracle}{semantic test oracle\xspace}

\renewcommand{\paragraph}[1]{\vskip 0.05in \noindent {\bf #1.}}

\def\hili{\leavevmode\rlap{\hbox to \hsize{\color{gray!20}\leaders\hrule height .8\baselineskip depth .5ex\hfill}}}

%% file: intro.tex
\section{Introduction}
Smart contracts are computer programs that execute on top of blockchains (e.g., Bitcoin~\cite{bitcoin} and Ethereum~\cite{ethereum}) to manage the flow of funds, exchange of assets, and transfer of digital rights between various parties~\cite{peters2016understanding,xu2017design}.
Transactions within smart contracts are stored persistently in the blockchain and thus immutable, without requiring a central third party to validate them.
Due to these unique advantages, smart contracts have gained a lot of popularity and 
attraction in recent years.
Many believe that this technology has the potential to reshape a number of industries, 
e.g., banking, insurance, supply chains, and financial exchanges~\cite{iansiti2017truth}.

The role of smart contracts in managing shared assets (often cryptographic currencies) requires a high level 
of security and reliability.
Yet, an increasing number of high-profile attacks have occurred, resulting in great financial 
losses.
Such attacks are facilitated by the lack of a rigorous development and testing process.
One notorious example is the ``DAO'' attack, where attackers stole more than $3.5$ million 
\ether (equivalent to about \$$45$ million \textit{USD} at that time) from ``DAO'' 
contract~\cite{chang2018scompile}.

These incidents have spurred activities in detecting vulnerabilities in smart 
contracts~\cite{chang2018scompile,tsankov2018securify,jiang2018contractfuzzer,luu2016making,nikolic2018finding,kalra2018zeus,liu2018reguard}.
Existing techniques usually detect smart contract vulnerabilities based on rule-based approaches: the contract behaviors are matched to a limited set of \emph{vulnerability patterns} identified beforehand.
The precision and recall of these techniques largely depend on the size and quality of 
their collections of vulnerability patterns.
Most such patterns are defined at the syntax level such as particular statements/calls sequences, 
ignoring their actual effects on contracts and resulting in false positives.
For example, Zeus~\cite{kalra2018zeus} treats the \texttt{refund} function in 
DaoChallenge~\cite{daoChallenge_source} contract (shown in Fig.~\ref{fig:reentrancyE}) as 
vulnerable, because it may potentially be reentered.
However, such reentrancy behavior cannot be exploited in stealing \ether from it, since 
the authors have incorporated defensive mechanisms to prevent from 
transferring unauthorized \ether (c.f. Sect.~\ref{sec:non-exploitable}).
Similar problems exist in how ContractFuzzer~\cite{jiang2018contractfuzzer} detects \emph{exception 
disorder} (more details in Sect.~\ref{sec:non-exploitable}). 

To address above issues, we propose to dynamically execute transactions and observe their 
actual effects on the contract states in order to detect exploitable vulnerabilities.
Our key insight is that almost all the existing (syntactical) vulnerability patterns result in a 
(semantic) mismatch between the externally visible events (e.\,g., amount transferred and contract 
balance changed) and the internal contract states (e.\,g., amount maintained in the contract's 
internal bookkeeping) in a transaction.
As a result, the internal bookkeeping becomes inconsistent, indicating a successful exploit.
Based on this observation, we define a general-purpose \SemanticTestOracle, which can be 
used to detect such mismatches at runtime.

Our technique is dynamic and works on target contracts that run on a realistic test environment.
Thus, it does not suffer from the imprecision faced by most static techniques.
All the vulnerabilities detected by our approach can be successfully reproduced.
When generating attack inputs, we take into account the unique characteristics of smart contracts  
which make traditional fuzzers ineffective. 
For example, attackers need to synthesize a sequence of transactions to successfully mount 
an attack (e.g., the transaction sequence ``\textit{deposit} $\rightarrow$ \textit{withdraw}'' 
is required for the DAO attack)~\cite{Feng2019PreciseAS}.
In contrast, traditional fuzzers such as AFL~\cite{zalewski2016american} focus on 
vulnerabilities triggered by a \emph{single} test case.
We extend traditional grey-box fuzzers with mutation operators customized for smart contracts, 
including \emph{transaction sequences}, \emph{gas limits}, \emph{fallback functions}, and 
\emph{contract states}, apart from \emph{function inputs}. 
We also develop the novel feedback mechanisms to guide the fuzzing process, by considering the data-flow and dynamic contract state information, together with the control-flow information.

We implemented our approach in \tool and evaluated it on 218 vulnerable smart 
contracts reported by
ContractFuzzer~\cite{jiang2018contractfuzzer} and Zeus~\cite{kalra2018zeus}.
We found that, of these potentially vulnerable contracts,
only 28 (12.84\%) are exploitable, and the remaining (87.16\%) are not.
In addition, \tool detected 26 hitherto unknown vulnerabilities, which could not be detected using 
previously identified vulnerability patterns.

In this paper, we make the following novel contributions:%
\begin{itemize}[leftmargin=*]
  \item We design a general-purpose \SemanticOracle, which can be used to detect a wide range of 
  vulnerabilities, such as reentrancy, exception disorder, gasless send 
  and integer overflow/underflow.
  \item We develop an oracle-supported dynamic exploit generation framework for smart contracts ---
  \tool. Specifically, we design customized mutation operators and feedback mechanisms, which are 
  shown useful at improving the effectiveness of vulnerability detection.
 \item We evaluate \tool on 218 smart contracts and demonstrate its advantages in 
 discovering \emph{exploitable} vulnerabilities over state-of-the-art techniques. 
  Among the 218 vulnerabilities reported by the state-of-the-art techniques, only 28 are exploitable, and \tool detects all of them without false positives.
  \item We present our findings on the 26 newly identified vulnerabilities, which cannot be detected by previously identified vulnerability patterns.
\end{itemize}

The rest of this paper is organized as follows. 
Sect.~\ref{sec:back} provides the necessary background and definitions for the rest of the paper.
Sect.~\ref{subsec:Oracle} illustrates our \SemanticTestOracle.
Sect.~\ref{sec:approach} introduces the technical details of our oracle-supported fuzzing and 
automated exploit generation.
Sect.~\ref{sec:eval} discusses challenges for implementing \tool and the evaluation results on real 
Ethereum smart contracts.
Finally, we discuss related work and conclude in Sect.~\ref{sec:relate} and 
Sect.~\ref{sec:conclude}, respectively.

%% file: background.tex
\section{Preliminaries}
\label{sec:back}

In this section, we provide the necessary background and definitions for the rest of the paper.

\subsection{Blockchain and Smart Contract}
A \emph{blockchain} is a shared, transparent distributed ledger, and is maintained by a 
decentralized network of peers (miners)~\cite{grech2018madmax}.
The miners perform the mining process of adding a block and verifying the validity of transactions 
through a \emph{proof-of-work} (PoW)~\cite{jakobsson1999proofs} or other consensus protocols, such 
as \emph{proof-of-stake} (PoS)~\cite{king2012ppcoin}.
Thus, a blockchain can be considered as an ever-growing list of blocks, each encoding a sequence of transactions, always available for inspection and safe from tampering.
Each block contains a cryptographic signature of its previous block.
No previous block can be changed or rejected, unless 51\% of miners are controlled and all its 
successors are changed or rejected.
With this structure, blockchain achieves decentralization, traceability, transparency, and 
immutability.

A \emph{smart contract} is computer program which allows users to define and execute transactions 
automatically on the blockchain~\cite{wu2018cream}.
A smart contract resides at a specific address on the blockchain, providing a number of publicly 
accessible functions and fields.
Moreover, a special \textit{balance} variable records the cryptocurrencies owned by the contract 
address and cannot be freely altered by programmers.
When a function of the smart contract is invoked, the current state of the contract is retrieved 
from the blockchain, and the updated state of the contract is stored back on the blockchain after 
execution.

A transaction is carried out in the form of a message sent to a particular address on the 
blockchain, which can either be a normal user account address or a contract address.
A user sends transactions to the blockchain in order to: (1) create new contracts, (2) invoke 
a function of a contract, or (3) transfer cryptocurrencies to contracts or other users.
All the transactions sent by participants, called \emph{external transactions}, are recorded on the 
blockchain.
Upon receiving an external transaction, a contract can also trigger some internal transactions, 
which are not explicitly recorded on the blockchain, but still have effects on the balance of 
participants or contracts.

The Ethereum Virtual Machine (EVM)~\cite{ethereum} is a stack-machine with an instruction set 
including standard arithmetic instructions, conditional and unconditional jump instructions, basic 
cryptography primitives, and primitives for gas computation.
The data is stored on the persistent memory area storage (a key-value store that maps 256-bit words 
to 256-bit words), the contract-local memory (a contract obtains a freshly cleared instance for 
each message call), or a stack (since the EVM is not a register machine but a stack machine, all 
computations are performed on the stack).
When Ethereum smart contracts are compiled and deployed, they are run on the EVM.

\begin{figure}[t]
  \centering
  \includegraphics[scale=.5]{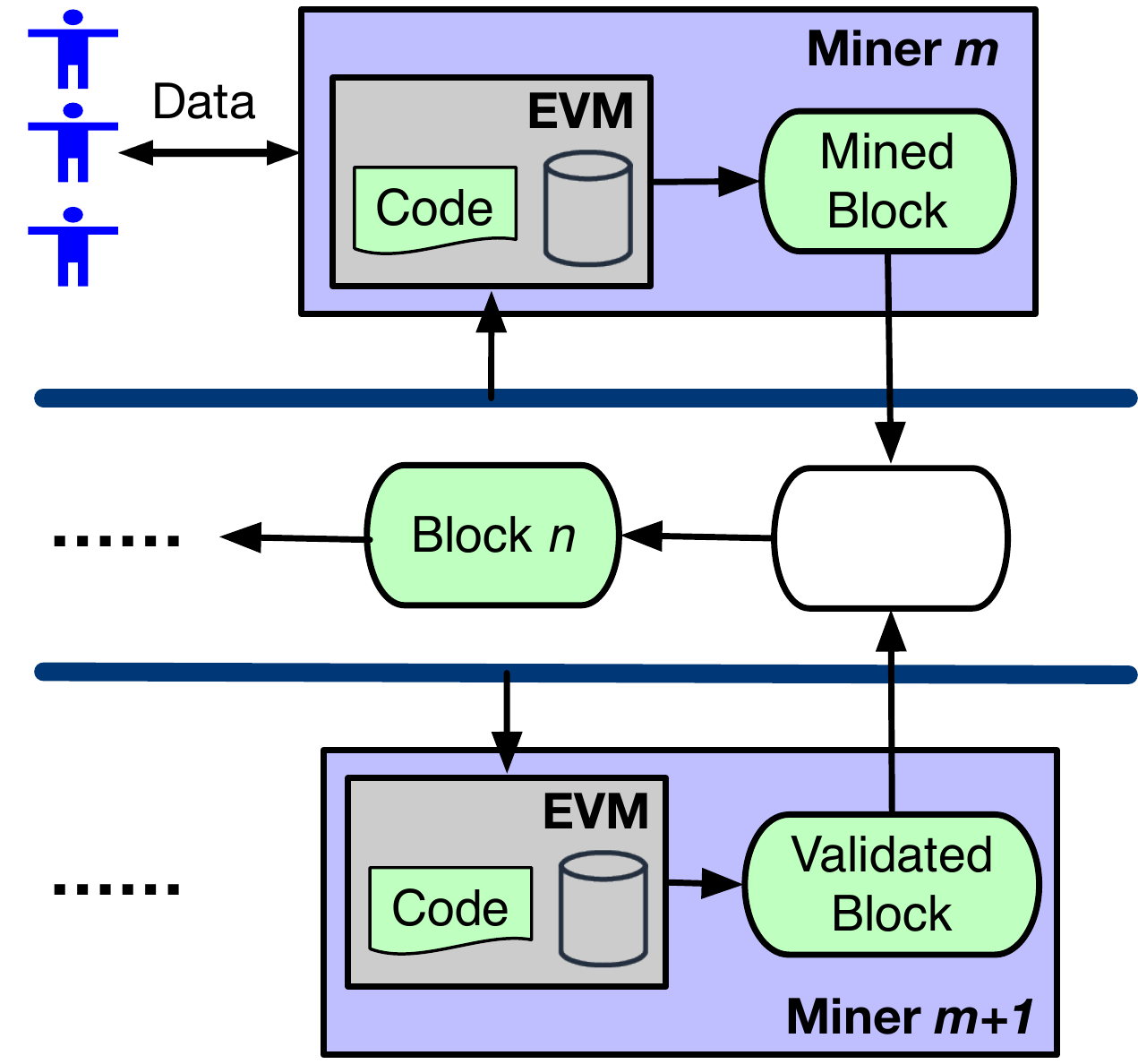}
  \caption{Illustration of Ethereum smart contracts running on blockchains.}
  \label{fig:bcworkflow}
\end{figure}

Fig.~\ref{fig:bcworkflow} is a schematic representation of the Ethereum smart contracts running on 
blockchains.
When participants submit a transaction to the blockchain, they may be mined by a \emph{miner} $m$ 
and executed on $m$'s EVM.
When the transaction is finished, the execution result is appended to the block list and propagated 
over the blockchain network.
At this point, other miners, such as $m+1$ may discover the new block list and validate it.
When majority of the miners validate the new block list, it becomes the confirmed block list, and 
cannot be changed or rejected.

\subsection{A Customized Semantic Model}
Smart contracts are similar to general computer programs in the sense that they are written in 
a Turing-complete language, e.\,g., Solidity~\cite{solidity}.
Various formal semantic models of Solidity
have been proposed in previous work~\cite{kalra2018zeus,jiao2018executable}, to enable formal 
verification of smart contracts.
Since our goal in this paper is to perform dynamic analysis for Solidity contracts and verify 
runtime contract states against our test oracle, we propose a customized semantic model
that is simple but expressive enough to capture the \emph{transaction-related} contract behaviors.
We define \emph{contract state} and \emph{transaction} as follows.


\begin{definition}[Contract and Contract State] \label{def:state}
  A \emph{contract} can be abstracted as a tuple $C:=\left\langle c, \mathit{bal}, A, \sigma 
  \right\rangle$, 
  where $c \in \mathit{Addr}$ is a unique address identifying the contract, $\mathit{bal} \in 
  \mathbb{N}$ is the externally visible \emph{balance} of the contract, $A \in 2^\mathit{Addr}$ is 
  a set of \emph{account addresses} of participants, and $\sigma$ is the internal contract state.
  A \emph{contract state} $\sigma$ is a type-consistent valuation of the global variables ($V$).
\end{definition}

The set of all states is denoted by $\Sigma \cup \{\mathit{Err}\}$, where $\mathit{Err}$ is a 
special state indicating an \emph{error state}.
For a given state $\sigma \in \Sigma$ and an expression $e$, $e_{\sigma}$ denotes the 
evaluation of $e$ in that state.
The semantics of a contract program is a set of \emph{execution traces}, where a trace 
corresponds to a sequence of internal contract states.
In Ethereum, each contract also has an externally-visible \emph{contract balance} 
($\mathit{bal}$) representing the total amount of funds in the contract, which is a part of the 
\emph{blockchain state} (as opposed to the contract state).

\begin{definition}[Transaction]
  A \emph{transaction} $t := \left\langle s, r, v \right\rangle$, if performed successfully, 
  deducts amount $v$ from the sending account's balance ($s.\mathit{bal}$ where $s \in \mathit{Addr}$) and transfers 
  the funds to a receiving account at address $r \in \mathit{Addr}$.
  We denote the values of a variable $g$ before and after the transaction as $pre(g)$ and 
  $post(g)$, respectively. 
\end{definition}

A transaction usually alters the blockchain state, reflected as updates on the contract balance.
In this case, the caller and callee contracts of a transaction correspond to the sending and 
receiving accounts, respectively.

\subsection{Threat Model}
To study the potential vulnerabilities of smart contracts, an important guarantee of a securely 
implemented contract is that it only allows \textit{authorized} accounts to transfer the 
\textit{authorized} amount of \ether~\cite{krupp2018teether}.
For example, the DAO contract, if implemented correctly, should only allow users to withdraw the 
amount of \ether have been deposited previously.
We assume that a regular user with no special capabilities attempts to break through this guarantee.
A smart contract is \emph{vulnerable}, if adversaries can bypass authorization and steal more 
\ether from the contract than allowed, resulting in a loss for contract owner, or victims store 
\ether in the contract but receive a lower level of authorization than intended, resulting in a 
loss for the participants~\cite{krupp2018teether}.

In this paper, we consider smart contracts which use internal bookkeeping variables to honestly
record the authorized amount for authorized participants.
This is common practice for contracts which manage shared funds.
For example, in the standard Ethereum contracts such as ERC-20~\cite{ERC20} and 
ERC-721~\cite{ERC721}, the function \texttt{balanceOf} returns the authorized amount for an 
authorized participant.

%% file: oracle.tex
\section{Semantic Test Oracle}\label{subsec:Oracle}
The fundamental difficulty in detecting smart contract vulnerabilities is the lack of a general-purpose test oracle.
This is because smart contracts do not crash like general computer programs, and their
execution may be silently reverted in cases of irregularities.
To address this issue, we propose a test oracle which detects irregularities in smart contracts at 
the semantic-level.
Our \SemanticTestOracle implements two types of invariants that transactions must comply with.
 %

\subsection{Balance and Transaction Invariants}
Smart contracts are mainly used to manage the transfer of assets and perform 
bookkeeping~\cite{chatterjee2018quantitative}, thus they need to keep track of participants' individual account balances, called the 
\emph{bookkeeping balances}.
Contract programs use an internal \emph{bookkeeping variable} (e.\,g., \texttt{balances} in 
Fig.~\ref{fig:dao}) to record the bookkeeping balances.
Suppose a bookkeeping variable $m : Addr \mapsto \mathbb{N}$ is given.

\begin{definition}[Balance Invariant] \label{def:BalanceInvariant}
	For every contract $\left\langle c,\allowbreak \mathit{bal},\allowbreak A,\allowbreak \sigma 
	\right\rangle$, $\sum_{a \in A} 
	m_{\sigma}(a) - \mathit{bal} = K$, where $K$ is a constant.
\end{definition}

The \emph{balance invariant} requires that the difference between the contract balance and the
sum of all participants' bookkeeping balances remains constant, before and after a transaction.
This invariant is defined within a single contract, i.\,e., \emph{intra-contract}, and it ensures 
the integrity of the bookkeeping balances.
If the bookkeeping balances are not updated correctly within a transaction, then the violation of this invariant indicates that an irregular event has happened.
For example, when an integer underflow happens during a transaction, the contract balance 
naturally goes down while the bookkeeping balances go up instead.

\begin{definition}[Transaction Invariant] \label{def:TransactionInvariant}
	For every 
	outgoing transaction $\left\langle c, r, v \right\rangle$ {where $c$ is the sending contract's}
	address, $\Delta(m_{\sigma}(r)) + 
	\Delta(r.\mathit{bal}) = 0$, where $\Delta(x) = \mathrm{post}(x) - \mathrm{pre}(x)$. 
\end{definition}

The \emph{transaction invariant} requires that the amount deducted from a contract's bookkeeping 
balances is always deposited into the recipient's balance.
This \emph{inter-contract} invariant ensures the consistency between the both ends of a transaction.
Note that the consistency of incoming transactions can be guaranteed by the balance invariant or 
other contract's outgoing transaction invariant.
In some cases, a transaction in progress may fail and funds are not transferred.
If the failure is not be captured by the contract's bookkeeping variables, the
contract may become vulnerable (e.\,g., exception disorder).

\subsection{Runtime Invariant Checking}
Since the invariants are supposed to hold for each transaction among contracts involved, the test oracle can be implemented as a set of runtime checks before and after each transaction. Notice that the runtime checks mentioned here are on the transaction-level, in which multiple contracts are involved. Thus, adding assertions in contracts does not work since assertions in each contract cannot express inter-contract properties.
The biggest challenge of implementing such runtime checks is to automatically identify the bookkeeping variables.


\paragraph{Identification of Bookkeeping Variables}
Most contracts performing meaningful transactions among multiple participants contain a 
bookkeeping variable, usually with the name \texttt{balances} or \texttt{balanceOf}.
The bookkeeping variable has a few characteristics distinguishing it from the others:
(1) it is a mapping from account addresses to unsigned integers, i.e.,
\texttt{mapping(address => uint*)} (there are a few exceptions which are explained in 
Sect.~\ref{sec:Evaluation});
(2) it is at least updated once in a \texttt{payable} function; and 
(3) in a normal transaction, the amount received from an account address should be reflected as a 
balance increase for that address.

Based on these observations, we design an algorithm for the automatic identification of bookkeeping 
variables.
For every mapping variable updated in \texttt{payable} functions, we send several transactions with 
randomly chosen values (including extremely large and small amounts).
We then observe the increased amount at the sender's address.
If the increases always match with the amounts being sent, we record the variable as a bookkeeping 
variable.

The bookkeeping variables in some contracts may not refer to the amount of \ether.
This is often the case in ERC-20 and ERC-721 contracts~\cite{ERC20}.
In these contracts, participants' digital assets are reflected in terms of the number of available 
\emph{tokens} rather than \ether.
In such cases, standard APIs for getting individual account balances (\texttt{balanceOf}) and total 
contract balances (\texttt{totalSupply}) in terms of tokens are provided and can be directly used 
to implement the runtime checks.

%

\subsection{Detecting Vulnerabilities with the Test Oracle} \label{sec:vulnerability}
Now we discuss how previously reported 
vulnerabilities~\cite{tsankov2018securify,nikolic2018finding,atzei2017survey,grech2018madmax} can 
be detected by our test oracle~\cite{wang2019vultron}.

\begin{figure}[t]
	\centering
	\includegraphics[width=.85\columnwidth]{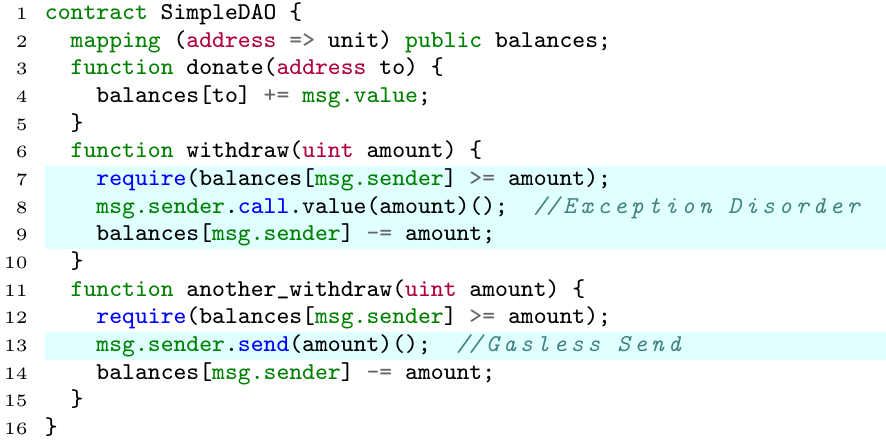}
	\caption{A simple contract susceptible to the ``DAO'' attack.}
	\label{fig:dao}
\end{figure}

\begin{figure}[t]
	\centering
	\includegraphics[width=.85\columnwidth]{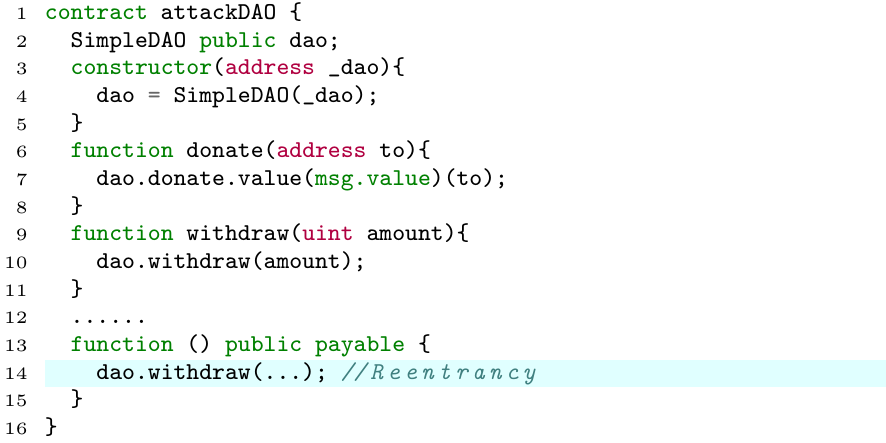}
	\caption{Reentrancy attack on the simple ``DAO'' contract.}
	\label{fig:reentrancy}
\end{figure}

\begin{figure}[t]
	\centering
	\includegraphics[width=.85\columnwidth]{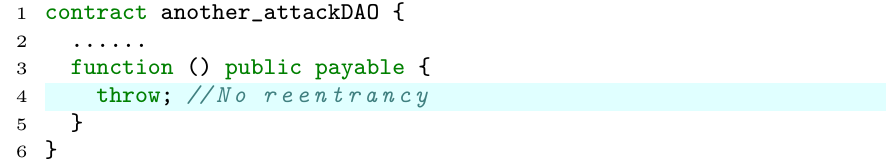}
	\caption{Exception disorder example.}
	\label{fig:exception}
\end{figure}

\paragraph{Reentrancy}
Programmers often believe that, when a non-recursive function is invoked, it cannot be re-entered 
before its termination.
However, this is not always the case, due to the fallback function introduced by Solidity.
Take the simplified ``DAO'' attack for example.
Two contracts, \texttt{SimpleDAO} (the victim, in Fig.~\ref{fig:dao}) and \texttt{attackDAO} (in Fig.~\ref{fig:reentrancy}), are deployed on the blockchain.
The reentrancy vulnerability of \texttt{SimpleDAO} can be exploited by 
\texttt{attackDAO}.
When \texttt{attackDAO} withdraws from \texttt{SimpleDAO} via Line 10 of Fig.~\ref{fig:reentrancy}, it will execute Lines 7--8 of Fig.~\ref{fig:dao}.
Then, due to fallback function mechanism, Line 8 of Fig.~\ref{fig:dao} executes Line 14 of Fig.~\ref{fig:reentrancy}, which further executes Lines 7--8 of Fig.~\ref{fig:dao} again and thus generate recursive calls.
Notice that, the execution of Line 9 of Fig.~\ref{fig:dao} is delayed.

The consequence of reentrancy is that Line 9 of Fig.~\ref{fig:dao} may be executed more times than allowed, and it leads to the integer underflow of bookkeeping variable \texttt{balances}.
The underflow will  produce the incorrect values for \texttt{balances}, which violates the balance invariant (Definition~\ref{def:BalanceInvariant}).

\paragraph{Exception Disorder}
Solidity is not uniform in handling exceptions.
Within a chain of nested calls, there are two types of exception handling 
mechanisms~\cite{atzei2017survey}:
(1) If a function in the chain is a \texttt{call} (the same for \texttt{delegatecall} and \texttt{send}), the exception is propagated along the chain, reverting all side effects, until it 
reaches the \texttt{call}.
From that point on, the execution is resumed with the \texttt{call} returning \texttt{false}.
(2) If all the functions in the chain are direct calls (not via \texttt{call}, \texttt{delegatecall} and \texttt{send}), the execution stops and all side effects are reverted, including the transfers of \ether.

Developers may handle exceptions incorrectly.
For example, Line 8 of Fig.~\ref{fig:dao} tries to transfer \ether to account \texttt{msg.sender}.
If this account is a contract, this transfer may fail, resulting in an exception.
Since this exception is not properly handled, the balance of this account
(\texttt{balances[msg.sender])} is decreased but it does not actually receive
\ether. Thus, this transaction will violate the transaction invariant (Definition~\ref{def:TransactionInvariant}).

\paragraph{Gasless Send}
\emph{Gasless send} is a special case of exception disorder.
When transferring \ether from one contract to another with function
\texttt{send}, it may lead to an out-of-gas exception.
The default gas limit for function \texttt{send} is 2,300~\wei.
If the recipient's fallback function contains too many instructions,
it may lead to an out-of-gas exception for function \texttt{send},
resulting in a \emph{gasless send}.
For example, at Line 13 of Fig.~\ref{fig:dao} the attacker (whose address is \texttt{msg.sender}) may have an expensive fallback function and the \texttt{send} function may fail. 
Since \emph{gasless send} is a special case of exception disorder, it can also be detected by the transaction invariant (Definition~\ref{def:TransactionInvariant}).

\begin{figure}[t]
	\centering
	\includegraphics[width=.85\columnwidth]{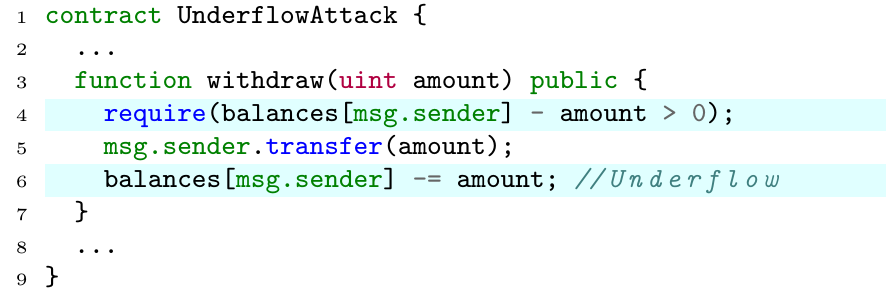}
	\caption{The underflow attack example~\cite{underflow}.}
	\label{fig:underflow}
\end{figure}

\paragraph{Integer Over/Under-flow}
Smart contracts heavily use integer arithmetic operations to manipulate participants' balances.
However, these variables are
susceptible to integer over/under-flow, e.g., in Fig.~\ref{fig:underflow}, \texttt{balances[msg.sender]} and 
\texttt{amount} are both unsigned integers.
If \texttt{balances[msg.sender]} is less than \texttt{amount}, the check at Line $4$ will pass due to integer underflow, leading to another underflow at Line $6$.
This produces the wrong value for the bookkeeping variable, which violates the balance invariant (Definition~\ref{def:BalanceInvariant}).

\paragraph{Other Vulnerabilities}
There are a few other types of vulnerabilities, including the \emph{timestamp 
dependency}, \emph{block number dependency} and \emph{freezing ether}~\cite{atzei2017survey}.
Exploiting vulnerabilities such as timestamp and block number dependencies requires the cooperation 
of miners, therefore cannot be easily realized at the contract-level.
In model checking~\cite{clarke2018model} terms, freezing ether is an violation of the 
\emph{liveness property}, while our dynamic approach can only detect violations of \emph{safety 
	properties}.
\tool mainly focuses on detecting reentrancy, exception disorder, gasless send, integer 
over/under-flow, and potentially other types of vulnerabilities triggered during transactions.

%% file: fuzzing.tex
\section{Oracle-Supported Fuzzing}
\label{sec:approach}

\begin{figure}[t]
	\centering
	\includegraphics[width=\columnwidth]{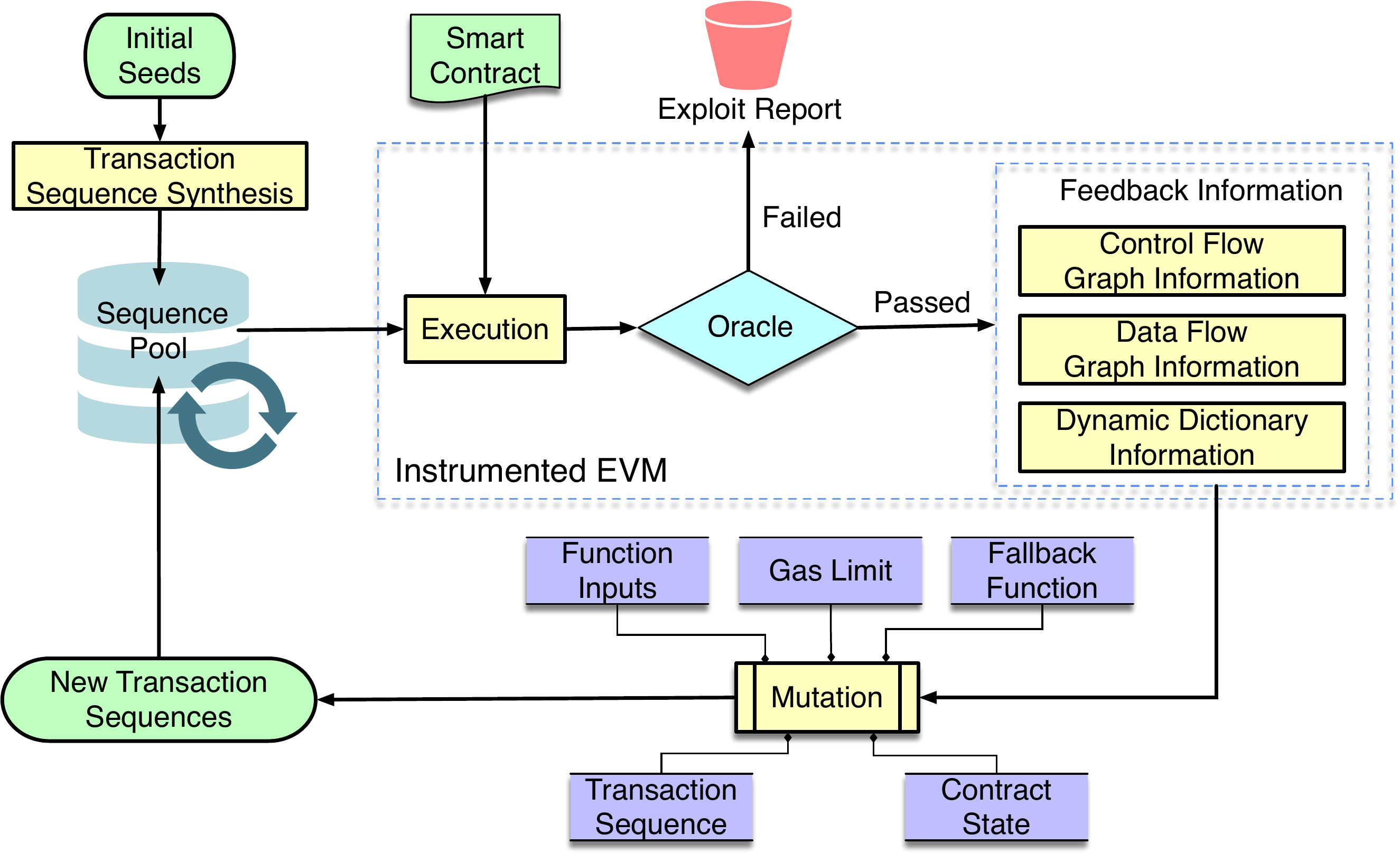}
	\caption{Overview of the \tool.}\label{fig:workflow}
\end{figure}

Fig.~\ref{fig:workflow} shows the overview of \tool, which is driven by a grey-box fuzzing 
loop~\cite{zalewski2016american,listeelix}.
Given a set of initial seeds, \tool randomly synthesizes a set of transaction sequences and picks one from 
the pool in each iteration of the fuzzing loop.
\tool runs each transaction in sequence on an instrumented EVM.
When the transaction is finished, \tool verifies the contract state against our 
\emph{\SemanticTestOracle} (details in Sect.~\ref{subsec:Oracle}).
If a violation is detected, \tool reports a vulnerability and presents the generated attack 
contract and transaction sequence which can be used to reproduce the exploit.
Otherwise, \tool collects runtime execution information to guide the test sequence generation for 
the next iteration.
The information collected mainly includes control-,  data-flow graph and contract state information.


Furthermore, \tool is equipped with a number of new mutation operators customized for smart 
contracts (c.\,f.~Sect.~\ref{subsec:Mutation}).
In addition to \emph{function inputs} used in traditional fuzzers, we also use 
\emph{gas limits}, \emph{fallback functions}, \emph{transaction sequences} and \emph{contract 
  states} as the mutation targets.
Many vulnerabilities in smart contracts require the interplay of several contracts and can only be 
exploited by a particular transaction sequence with the correct gas limit~\cite{kolluri2019exploiting}.
Therefore, the customized mutation operators are important for triggering vulnerabilities.
The newly generated inputs are added to the pool and the fuzzing process continues until it exceeds 
the allocated resource limits.

\subsection{Oracle-Supported Fuzzing Algorithm}\label{subsec:Mutation}
The goal of the fuzzing component is to automatically generate transaction sequences that violate 
the test oracle.
Algorithm~\ref{alg:fuzzing} presents its high-level idea.
Given a contract program $C$ and a set of initial seeds $T$, it first generates a set of initial 
transaction sequences $\mathit{TS}$ from $T$ (Line~\ref{fuzzing:initial}).
In every iteration of the fuzzing loop (Lines~\ref{fuzzing:loophead}--\ref{fuzzing:looptail}),
we select a transaction sequence $\mathit{ts}$ from $\mathit{TS}$ (Line~\ref{fuzzing:select}), and 
initializes the current execution trace sequence ($E$) and contract state dictionary ($dict$) 
as $empty$ (Line~\ref{fuzzing:Einitial}).
\tool then executes each transaction $t_i \in \mathit{ts}$
(Lines~\ref{fuzzing:Eloophead}--\ref{fuzzing:Elooptail}),
and collects its execution trace $e_i$ (Line~\ref{fuzzing:run}).
It would also stores observed contract states (e.g., the values of bookkeeping variables and 
contract balances) into dictionary (Line~\ref{fuzzing:extract}), similar to what is done in 
traditional fuzzers such as VUzzer~\cite{rawat2017vuzzer}.
The dictionary values are later used in generating function inputs.
The execution trace of each transaction $e_i$ is concatenated to form a transaction sequence trace 
$E$ (Line~\ref{fuzzing:append}).
After that, \tool checks whether the test oracle is violated.
If so, it adds the current transaction sequence $\mathit{ts}$ into the output $\mathit{TS}'$, which 
is a script exploiting the vulnerability (Line~\ref{fuzzing:add}).

\tool performs mutations at both the level of single transaction 
(Lines~\ref{fuzzing:Mloophead}--\ref{fuzzing:Mlooptail}), and at the level of transaction sequences
(Lines~\ref{fuzzing:MTranshead}--\ref{fuzzing:MTranstail}).
If $t_i$ achieves new branch coverage, we perform mutations on the function inputs and gas limit 
(Lines~\ref{fuzzing:input}--\ref{fuzzing:gas}).
When fallback function is called in the transaction, we also perform mutations on fallback 
functions at Line~\ref{fuzzing:fallback}.
If the transaction sequence's trace $E$ has new data dependence coverage, we perform transaction 
sequence mutation (e.g., switching the order of transactions) at Line~\ref{fuzzing:tranSequence}.
In the end, we randomly reset the whole smart contract states by contract state mutation at 
Line~\ref{fuzzing:tranState}.
More details about each mutation strategy are given in Sect.~\ref{sec:mutation-strategy}.

\LinesNumbered
\begin{algorithm}[tb]
	\small
	\SetKwInOut{Input}{input}
	\SetKwInOut{Output}{output} 
	\SetKwComment{Comment}{$\quad //$}{}
	\Input{a contract program $C$, a set of initial seeds $T$
	}
	\Output{transaction sequences $\mathit{TS}'$ violating the test oracle}
	\BlankLine
	$\mathit{TS} \leftarrow$ generate initial transaction sequences from $\textit{T}$ \; \label{fuzzing:initial}
	\While{time budget not reached and abort signal not received}{ \label{fuzzing:loophead}
		$\mathit{ts} \leftarrow \textit{selectNext(TS)}$ \;  \label{fuzzing:select}
		$E, dict \leftarrow \emptyset, \emptyset$ \;  \label{fuzzing:Einitial}
		\hili \ForEach{transaction $t_i$ in $\mathit{ts}$}{ \label{fuzzing:Eloophead}
			\hili run $t_i$ and collect execution trace $e_i$ on \textit{EVM} \;  \label{fuzzing:run}
			\hili $dict \gets \textit{extractValues}(e_i)$ (store dictionary values) \;
			\label{fuzzing:extract}
			\hili $E \gets E, e_i$ (append $e_i$ to $E$) \; \label{fuzzing:append}
			\If{test oracle is violated}{
				$\mathit{TS}' \leftarrow \mathit{TS}' \cup \mathit{ts}$ \; \label{fuzzing:Elooptail} \label{fuzzing:add}
			}
		}
		\ForEach{transaction $t_i$ in $\mathit{ts}$}{   \label{fuzzing:Mloophead}
			\If{$t_i$ has new branch coverage in $e_i$}{
				$\mathit{TS}_i \leftarrow$ InputMutate($t_i$, $dict$) \;   \label{fuzzing:input}
				$\mathit{TS}_g \leftarrow$ GasMutate($t_i$) \;    \label{fuzzing:gas}
				$\mathit{TS} \leftarrow \mathit{TS} \cup \mathit{TS}_i \cup \mathit{TS}_g$ \;
				\hili\If{$t_i$ executes fallback function}{  \label{fuzzing:fallbackhead}
					\hili$\mathit{TS}_f \leftarrow$ FallbackMutate($t_i$) \;  \label{fuzzing:fallback}
					\hili$\mathit{TS} \leftarrow \mathit{TS} \cup \mathit{TS}_f$ \;   \label{fuzzing:Mlooptail} \label{fuzzing:fallbacktail}
				}
			}	
		}	
		\hili\If{$E$ has new data dependence coverage}{  \label{fuzzing:MTranshead}
			\hili$\mathit{TS}_t \leftarrow$ TransSeqenceMutate($\mathit{ts},E,dict$) \; 
			\label{fuzzing:tranSequence}
			\hili$\mathit{TS} \leftarrow \mathit{TS} \cup \mathit{TS}_t$ \;
		} 
		\hili ContractStateMutate() \;  \label{fuzzing:looptail}   \label{fuzzing:MTranstail} \label{fuzzing:tranState}
	}
	\caption{Oracle-Supported Fuzzing}
	\label{alg:fuzzing}
\end{algorithm}

The highlighted code in Algorithm~\ref{alg:fuzzing} shows the differences of \tool from traditional 
grey-box fuzzing approaches such as AFL~\cite{zalewski2016american}.
To summarize, traditional fuzzing techniques work on a single call, while \tool works on a transaction sequence.
The reason is that a lot of vulnerabilities can only be triggered by a sequence of transactions.
To effectively generate such sequences, we use data-flow information to guide its mutation, which cannot be achieved by control-flow information.
Another important difference is that \tool performs mutations on fallback functions, through which the attack contracts may interact with the target contract.

\subsection{Mutation Strategies} \label{sec:mutation-strategy}
In this section, we present our five mutation strategies, namely, the mutation of 
\emph{function inputs}, \emph{gas limit}, \emph{fallback function}, \emph{transaction sequence}, 
and 
\emph{contract state}.

\paragraph{Attack Contract}
\tool uses the attack contracts to interact with target contract, thus we first automatically generate the attack contract, like in the example shown in 
Fig.~\ref{fig:reentrancy}.
To synthesize the attack contracts, we use a variable to represent
the target contract and initialize it in the constructor function
(lines 2--5).
Then, for each function in target contract, \tool develops a surrogate function to call this 
function, as shown in lines 5--11 in Fig.~\ref{fig:reentrancy}.
Finally, we synthesize the fallback function as shown in lines 12--14.

\paragraph{Function Inputs}
Line~\ref{fuzzing:input} of Algorithm~\ref{alg:fuzzing} mutates the parameters passed to each target function.
We consider two types of function parameters: primitive and array types.

Primitive-type parameters include Booleans (\texttt{bool}), account addresses, unsigned integers 
(\texttt{uint*}), integers (\texttt{int*}), and arbitrary-length raw byte data (\texttt{byte*}).
First, we pick special values from the dynamic dictionary of previously seen state variable values with 
matching types to generate multiple \emph{mutation ranges}.
Within these ranges, we randomly generate values as candidate function inputs.
Second, we opportunistically negate bits in these inputs to produce new inputs (similar to the 
``flip1'' operation used in AFL~\cite{zalewski2016american}).
For account addresses, we simply enumerate addresses from a predefined account list.
In most cases, the collected dynamic dictionary and ``flip1'' are enough for generating effective inputs, since most smart contracts have 
relatively simple program logic.

For array types, we consider both fixed- and arbitrary-length arrays.
For fixed-length arrays of primitive-type elements (e.\,g., \texttt{address[n]} and 
\texttt{uint*[n]}), we use the same technique described above to generate random values for each 
element.
For an arbitrary-length array, we first generate a positive random number as the array length, and 
then proceed as dealing with a fixed-length array.
For arbitrary-length bytes or strings, we use values from the dictionary and mutate them with bit 
flips.

\paragraph{Gas Limit}
Every instruction executed on EVM has an associated fee, known as the \emph{gas}.
If the gas cost of a transaction exceeds the \emph{gas limit}, an out-of-gas exception is thrown.
To simulate all possible behaviors with the exceptions thrown in the middle of a transaction, we 
mutate on the gas limits at Line~\ref{fuzzing:gas} of Algorithm~\ref{alg:fuzzing}.
First, we estimate the maximum gas cost $G_t$ and the intrinsic gas cost $G_i$~\cite{wood2018ethereum} 
(consisting of a constant transaction fee and a data-dependent fee) for a target transaction.
Then, we divide the range between $G_i$ and $G_t$ into $n$ equal intervals and randomly choose a 
gas limit from each interval to initiate the transaction with.

\paragraph{Fallback Function}
The fallback function is an important mechanism in Ethereum and is highly relevant to the 
reentrancy and exception disorder vulnerabilities.
When receiving funds from the target contract under test, the attacker's contract may use specially 
crafted fallback function to perform malicious activities.
To trigger these behaviors, \tool performs mutations on the fallback function at Line~\ref{fuzzing:fallback} of Algorithm~\ref{alg:fuzzing}.

We generate multiple attack contracts with different fallback 
functions to interact with the target contract, such as in Figs.~\ref{fig:reentrancy} and~\ref{fig:exception}.
In particular, we allow any function of the target contract to be called within the attacker's fallback function.
Apart from that, we also have an empty fallback function and one that contains a single 
\texttt{throw} statement (e.g., \texttt{revert}()) to trigger exception 
disorders.

\paragraph{Transaction Sequence}
Some vulnerabilities can only be triggered with the correct transaction sequences.
For example, the DAO attack can only be mounted by first depositing into the target contract and 
then withdrawing from it.
To find a successful exploit, we mutate the call sequences as follows.
For a given candidate transaction sequence, (1) if two transactions of a sequence
operate on the same contract state variable, we switch their order; (2) we
randomly select a transaction in the sequence, and replace it with a random new transaction;
(3) we randomly select a transaction in the sequence, and delete it;
(4) we randomly select a transaction in the sequence, and insert a random new transaction
before it.

\paragraph{Contract State}
The effects of a transaction depend on the contract state in which it is initiated.
To mutate contract states, we allow the values of state variables to be carried forward across 
multiple test runs and reset the state periodically, say, after every $n$ transactions.
The reset of contract state is achieved by redeploying the contract code to the private test 
network.

\subsection{Feedback Mechanisms} \label{subsec:feedback}
The feedback used by \tool can be broadly categorized into the \emph{control-driven} and 
\emph{data-driven}, and \emph{contract state} feedback information.
The control-driven feedback mechanism strives to cover more CFG edges as with AFL~\cite{zalewski2016american}, by favoring uncovered CFG edges.

%

\paragraph{Data-Driven Feedback}
Since smart contract is state-relevant, we should synthesize a suitable transaction sequence to detect the vulnerabilities.
However, the transaction sequence cannot be guided by the control-flow
information, as a different transaction sequence does not necessarily cover new CFG edges.
Thus, we propose to use data flow to guide transaction sequence mutations.
If the mutated transaction sequence covers new data dependencies, it is an
interesting transaction sequence.
We first define the data dependency as follows.

\begin{definition}[Data Dependency~\cite{cfg,data-dependency}]\label{def:dataD}
There is a data dependency from $y$ to $x$ if there exists a directed path $p$ from $x$ to $y$ 
where $x$ defines a variable $v \in V$, $y$ uses $v$ and there is no node $z \in p$ that redefines $v$.
\end{definition}

In the execution of transaction sequence, if two transactions operate on the same contract state variable, we switch their order to generate new transaction sequence.
For example, the transaction sequence ``\textit{withdraw} $\rightarrow$ \textit{deposit}'' both operate on the bookkeeping variable \textit{balances}, thus we switch their order to generate a new transaction sequence ``\textit{deposit} $\rightarrow$ \textit{withdraw}'', which may trigger the reentrancy vulnerability.

\paragraph{Contract State Feedback}
Apart from the data dependency, we also use the contract states 
(Definition~\ref{def:state}) to guide the function input generation.
The basic idea is that, in most cases, the execution of current transaction heavily depends on the contracts' states.
For example, the sequence ``\textit{deposit} $\rightarrow$ \textit{withdraw}'' may trigger the reentrancy, but it depends on whether the funds deposited is greater than the funds withdrawn.
Thus, we use contract states to guide function input generation, such that the less funds withdrawn than those deposited. 
to subsequently trigger potential vulnerabilities.
In fact, we extract the dynamic contract states as a dynamic dictionary,
which is similar to VUzzer~\cite{rawat2017vuzzer}. However, the latter uses the
\texttt{immediate} values in the code as the static dictionary.

\subsection{\tool by Example}
Take the DAO contract in Fig.~\ref{fig:dao} as an example.
Based on the initial seeds \{\textit{withdraw(10)}, \textit{deposit.value(5)(*)}\}, \tool randomly 
synthesizes a transaction sequence, ``$\mathit{ts}_1$ = \textit{withdraw(10)} $\rightarrow$ 
\textit{deposit.value(5)(*)}''.
After $\mathit{ts}_1$ is executed, we identify a data dependency between \textit{withdraw} and 
\textit{deposit} over the state variable \textit{balances} by analyzing the data flow of the 
execution trace.
Using this as feedback, we mutate $\mathit{ts}_1$ and generate ``$\mathit{ts}_2$ =  
\textit{deposit.value(5)(*)} $\rightarrow$ \textit{withdraw(10)}''. 
In $\mathit{ts}_2$'s execution, we assume it does not produce the reentrancy, because the value 
withdrawn is greater than deposited.
In addition, \tool extracts the contract states as the dynamic dictionary, e.g., the amount 
deposited into \textit{balances} is~$5$.
Next, \tool randomly generates three group of inputs ($\leq 5$, $5$, and $\geq 5$) as the 
parameters of \textit{withdraw}, e.g., we get ``$\mathit{ts}_3$ =  \textit{deposit.value(5)(*)} 
$\rightarrow$ \textit{withdraw(3)}''.
At present, we still assume the reentrancy does not happen, because the attack contract may not 
contain the right fallback function, e.g., \textit{another\_attackDAO} in Fig.~\ref{fig:exception}.
Since \tool also performs mutation on fallback functions, and thus may generate new fallback functions, as \textit{attackDAO} in Fig.~\ref{fig:reentrancy}.
Finally, we have \textit{attackDAO} in Fig.~\ref{fig:reentrancy}, as attack contract, and execute ``\textit{deposit.value(5)(*)} $\rightarrow$ \textit{withdraw(3)}'', which triggers reentrancy.

When reentrancy occurs, \ether would be transferred to the attacker's account via the 
\textit{withdraw} function. At last, the update of \textit{balances} at Line 9 would produce the underflow of balances.
Since the value of bookkeeping variable \textit{balances} is wrong, and thus the balance invariant (Definition~\ref{def:BalanceInvariant}) is violated.


%% file: eval.tex
\section{Implementation and Evaluation} \label{sec:Evaluation}
\label{sec:eval}

In this section, we discuss the implementation of \tool and evaluate it on a set of benchmarks.

\subsection{Implementation}
We use go-ethereum 1.8.20 to build a private Ethereum blockchain as the test network, and Truffle 
Suite 4.1.14 as the test harness. \tool consists of a front-end, which generates inputs and 
triggers transactions, and a back-end, which executes the smart contracts and validates its 
behavior.

The front-end triggers transactions and performs mutations based on feedback from earlier test runs.
In the back-end, the stock EVM is modified to monitor the runtime execution:
the test oracle is enforced by asserting invariants after each transaction is finished.
If an invariant violation happens, the test sequence is reported as an exploit.
Otherwise, it performs data-flow, control-flow, and contract-state analysis to provide feedback to 
the front-end, which continues to generate new test inputs.
In total, \tool is implemented with more than 5,000 lines of Javascript, Python and Go 
languages.

\subsection{Evaluation}
Our empirical evaluation of \tool tries to answer the following research questions:
\begin{itemize}[leftmargin=*]
    \item \textbf{RQ1:} How does \tool perform compared to the state-of-the-art pattern-based approaches?
    \item \textbf{RQ2:} How effective is the feedback-guided test generation in speeding up the exploit generation?
    \item \textbf{RQ3:} Can \tool discover previously unknown vulnerabilities?
\end{itemize}




\paragraph{Setup}
All our experiments were performed on a 64-bit Ubuntu~18.04 desktop with an Intel Xeon CPU E5-1650 (3.60\,GHz, 12 cores) and 16\,GB of RAM.
Since we focus on smart contract vulnerabilities, not the consensus protocol, we configure only one peer node for the mining process. 
We set the initial mining difficulty of the genesis block to $1$ so that transaction confirmation is fast.
We also assume that each participant owns as much \ether as the total \ether supply at the time of writing (currently about $10^8$\,\ether).

\paragraph{Subjects}
To evaluate our approach, we selected the experimental subjects as follows.
We compared \tool with ContractFuzzer~\cite{jiang2018contractfuzzer}, which is currently the only other dynamic fuzzing tool, and the static verification tool Zeus~\cite{kalra2018zeus}. 
These two tools reflect the state of the art in dynamic and static smart-contract analysis, and use properties that are stronger than our test oracle.
Thus, we use the reported vulnerabilities as our experimental subjects (except \emph{timestamp dependency}, \emph{block number dependency}, and \emph{dangerous delegatecall}, which cannot be easily exploited).

First, we included all the 188 contracts reported as vulnerable by ContractFuzzer into our benchmark.
Since ContractFuzzer does not analyze integer overflow/underflow~\cite{jiang2018contractfuzzer}, we 
augment the set of benchmarks with 30 contracts containing overflow/underflow vulnerabilities.
These 30 contracts were randomly selected from a set of 1,095 smart contracts that reported by 
Zeus~\cite{kalra2018zeus} to have this vulnerability.
In total, we selected $218$ smart contracts for our experiments, among which there are only $33$ ($15$\%) contracts whose bookkeeping variables cannot be automatically identified. We have manually investigated these contracts and found that they use complicated data structures such as structures or user-defined types to record the bookkeeping. Our simple heuristic is able to automatically identify bookkeeping variables for $85$\% of the contracts.

\subsection{Comparison with the State-of-the-Art} \label{subsec:Compare}
RQ~$1$ and RQ~$3$ relate to the effectiveness of \tool. To
evaluate this, we compared \tool with ContractFuzzer and Zeus. 
Table~\ref{tab:overall} shows the vulnerability type, the number of reported vulnerabilities, the number of actually exploitable vulnerabilities, and the percentage of exploitable vulnerabilities over the vulnerabilities reported by the state of the art. The last column lists the exploitable vulnerabilities reported by \tool.
Row ``\textit{New Vulnerabilities}'' shows the new vulnerabilities \tool found; these are different from the vulnerability types 
covered by the state-of-the-art ContractFuzzer and Zeus.


\begin{table}[t]
	\caption{Vulnerabilities reported by ContractFuzzer, Zeus and \tool.}
	\label{tab:overall}
  \centering
	\begin{tabular}{p{1.5in}rrrr}
		\toprule
		\multirow{2}{*}{Vulnerability Types} & \multicolumn{3}{c}{Pattern-based Detection} &
		\multirow{2}{*}{\tool} \\
		& \#Vul & \#Exp & (\%) & \\
		\midrule
		Reentrancy    & 14  &  6   &  42.86\,\% & 6     \\ 
		Exception Disorder   & 36 & 13  &  36.11\,\% & 13 \\ 
		Gasless Send    & 138  & 6  & 4.34\,\% &6   \\ 
		Integer Over/Under-flow    & 30 & 3 &  10.00\,\% & 3  \\      
		New Vulnerabilities   & --- & ---  &--- & 26 \\
		\midrule
		Total/Avg. & 218 & 28 & 12.84\,\% &  54 \\
		\bottomrule
	\end{tabular}
\end{table}


ContractFuzzer reports $14$ reentrancy vulnerabilities.
Out of these, only $6$ contracts were reported exploitable by \tool. For the $8$ smart contracts not reported by \tool, we manually checked the contract code and confirmed that they are non-exploitable.
Our investigation showed that ContractFuzzer over-reports reentrancy vulnerabilities (around 42.86\,\% exploitable) because its oracle is defined at the syntactic level. 

For the exception disorder, ContractFuzzer reported $36$ vulnerabilities, while \tool only reported $13$. 
ContractFuzzer only considers a transaction being safe if the exceptional case is followed by a \texttt{throw} statement.
However, an exception can be handled by multiple ways, e.\,g., reverting the modified variables, 
in which the exception would not lead to an exploitable vulnerability.

ContractFuzzer also reports 138 gasless send vulnerabilities.
However, the gasless send vulnerabilities are not exploitable if the \texttt{transfer()} function is used to send \ether,
because the \texttt{transfer()} function automatically reverts the program state if there is not enough gas.
These cases were reported by ContractFuzzer as vulnerable.
Out of the 138 gasless send vulnerabilities, only 6 were reported exploitable by \tool (4.34\,\%).

Integer overflow/underflows constitute another important issue in smart contracts.
Zeus detects integer overflow/underflow based on the predefined syntactic patterns~\cite{integeroverflow}.
However, whether these really happen depends on the execution environment and program contexts.
In the 30 sampled integer overflow/underflow vulnerabilities, only 3 were reported exploitable by \tool (10.00\,\%).

\paragraph{Summary} In the 218 detected vulnerabilities by ContractFuzzer and Zeus, only 28 vulnerabilities were reported exploitable by \tool (12.84\%). 
For the non-exploitable vulnerabilities, we manually checked and confirmed that they are indeed not exploitable. 
Furthermore, \tool finds 26 new vulnerabilities, which are different from the vulnerability types in ContractFuzzer and Zeus. 

From the above experiments, we observe that \tool only reports exploitable vulnerabilities because its oracle is defined
at the semantic level. In addition to that,
\tool is able to discover previously unknown vulnerabilities. Two of the authors have independently verified the results by replaying the exploit scripts manually. The authors of ContractFuzzer also confirmed our findings.
Later, in Section~\ref{sec:case-studies}, we explain why some vulnerabilities reported by ContractFuzzer and Zeus are not exploitable, and
illustrate the new attacks.



\subsection{Evaluation of Effectiveness}
RQ~$2$ questions the effectiveness of feedback in fuzzing.
To answer this question, we implemented a variant of \tool (called \random), which only uses the control-flow information to guide the fuzzing process, similar to AFL.
Then, we performed the experiments on the 23 exploitable examples (5 exploitable vulnerabilities are repetitive in \textit{exception disorder} and \textit{gasless send}) by repeating each experiment 8 times, and the compared the performance of \tool and \random. 
We set a timeout of 600 seconds for each benchmark program, in which \tool can successfully finish all experiments.

\begin{figure}[t]
	\centering
	\includegraphics[width=.7\columnwidth]{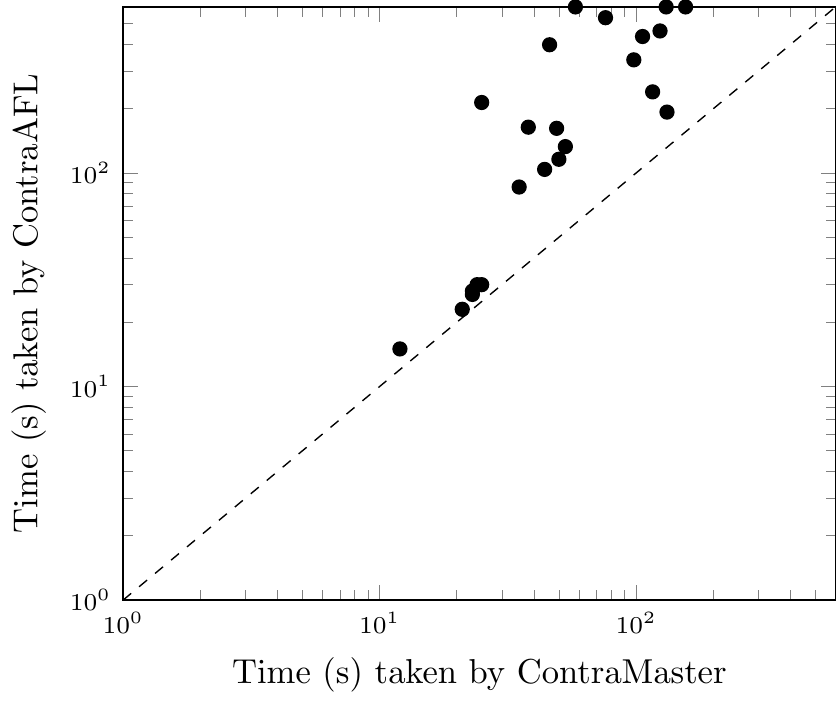}
	\caption{Time taken by \tool and \random.}\label{fig:compare}
\end{figure}

Fig.~\ref{fig:compare} shows the comparison results, where the $x$ and $y$ axes show the time taken by \tool and \random, respectively.
\tool performs better than \random for points above the diagonal line, which was observed for all examples we ran.
From the results, we can see that \tool is highly efficient, compared with \random, in recognizing exploitable vulnerabilities. 
Specially, 3 exploitable vulnerabilities cannot be found by \random in the given timeout (points lying on the top $x$-axis) in their all experiments.


\begin{table}[ht]%
	\centering
	\caption{Statistical results for feedback-directed fuzzing.}
	\label{tab:experiments}
\begin{tabular}{lrrrrrr}
	\toprule 
	\multirow{2}{*}{Contracts} & \multicolumn{2}{c}{
			 \tool(s)
	} & \multicolumn{2}{c}{\random(s)} & \multicolumn{2}{c}{Statistics}\tabularnewline
	\cmidrule{2-7} \cmidrule{3-7} \cmidrule{4-7} \cmidrule{5-7} \cmidrule{6-7} \cmidrule{7-7} 
	& Avg. & Variance & Avg. & Var & p-value & $\widehat{A}_{12}$\tabularnewline
	\midrule 
	BountyHunt & 25 & 137 & 214 & 62081 & 0.0043 & 0.898\tabularnewline
	Eth\_VAULT & 46 & 2148 & 399 & 57003 & 0.0035 & 0.906\tabularnewline
	PIGGY\_BANK & 98 & 5940 & 339 & 79022 & 0.0625 & 0.734\tabularnewline
	Private\_accumulation\_fund & 124 & 12614 & 463 & 64123 & 0.0135 & 0.828\tabularnewline
	Private\_Bank & 76 & 3412 & 534 & 34848 & 0.0013 & 0.937\tabularnewline
	PrivateDeposit & 156 & 8555 & 600 & 0 & 0.0002 & 1.000\tabularnewline
	EthSplit & 21 & 14 & 23 & 45 & 0.3952 & 0.547\tabularnewline
	FreeEth & 38 & 718 & 164 & 25189 & 0.0010 & 0.968\tabularnewline
	HelpMeSave & 50 & 1253 & 116 & 9234 & 0.0328 & 0.781\tabularnewline
	HFConditionalTransfer & 12 & 9 & 15 & 10 & 0.0755 & 0.719\tabularnewline
	Honey & 44 & 524 & 104 & 10249 & 0.0704 & 0.727\tabularnewline
	MultipicatorX4 & 35 & 401 & 86 & 3659 & 0.0091 & 0.859\tabularnewline
	Pie & 53 & 1224 & 133 & 5455 & 0.0203 & 0.813\tabularnewline
	TokenBank & 131 & 1066 & 600 & 0 & 0.0002 & 1.000\tabularnewline
	transferIntwopart & 23 & 49 & 27 & 119 & 0.2468 & 0.609\tabularnewline
	WahleGIveaway & 49 & 1153 & 162 & 26793 & 0.0781 & 0.718\tabularnewline
	CreditDepositBank & 106 & 4070 & 436 & 40511 & 0.0009 & 0.968\tabularnewline
	SafeConditionalHFTransfer & 25 & 37 & 30 & 75 & 0.1465 & 0.664\tabularnewline
	Soleau & 24 & 75 & 30 & 101 & 0.1125 & 0.688\tabularnewline
	Etheramid & 23 & 27 & 28 & 26793 & 0.0940 & 0.703\tabularnewline
	SImpleLotto & 132 & 879 & 193 & 2896 & 0.0157 & 0.828\tabularnewline
	MyToken & 116 & 651 & 240 & 25187 & 0.0023 & 0.929\tabularnewline
	SimpleCoinFlipGame & 58 & 90 & 600 & 0 & 0.0002 & 1.000\tabularnewline
	\midrule 
	Avg. & 63.58 & 1958.62 & 240.60 & 20582.36 & --- & ---\tabularnewline
	\bottomrule 
\end{tabular}
\end{table}

Furthermore, a manual investigation revealed that the test sequences generated by \random are mostly meaningless.
For example, \random often chooses amounts of \ether to send that are larger than the amount it owns.
Thus, the transaction would be reverted, resulting in no actual effect on the smart contracts.
On the other hand, \tool is guided by feedback and gradually generates meaningful transactions.

\paragraph{Mann Whitney U-test Scoring}
Following Klees et al.'s~\cite{klees2018evaluating} recommendation,
we apply the \emph{Mann Whitney U-test} on the time used to find the vulnerabilities.
As shown in Tab.~\ref{tab:experiments}, in most experiments,
the p-values are smaller than or close to a significance level of 0.05.
Thus, we conclude there exists a statistically significant difference in the time used to find the vulnerabilities, compared to \random.

\paragraph{Vargha and Delaney $\widehat{A}_{12}$ Scoring}
To determine the extent to which \tool outperforms \random,  we also use
{Vargha and Delaney's $\widehat{A}_{12}$}
statistical test~\cite{klees2018evaluating}.
From Tab.~\ref{tab:experiments}, we can see among benchmark experiments the
resulting $\widehat{A}_{12}$ statistic exceeds the conventionally large effect size of 0.71 in 18 out of 23 cases (78.3\,\%).
Therefore, we conclude that the time usage in \tool to find vulnerabilities is statistically different from that in \random.

%


\subsection{Case Studies}
In this section, we report on interesting findings from our case studies.
Sect.~\ref{sec:attack} introduces some new attacks found by \tool in the experiments, and 
Sect.~\ref{sec:non-exploitable} investigates on some non-exploitable vulnerabilities reported by 
ContractFuzzer~\cite{jiang2018contractfuzzer}, Zeus~\cite{kalra2018zeus}, and 
Oyente~\cite{luu2016making}.

\subsubsection{New Attack Surfaces}\label{sec:attack}
There are three different types of new attacks found by \tool in 26 smart contracts.


\paragraph{Incorrect Access Control}
Access control is important in smart contracts, which allows critical operations to be performed only by the owner of the contract.
Access control-related issues are ranked the second most severe among all vulnerability types~\cite{dasp}.
It is very hard to define a pattern/property to capture all incorrect access controls, unlike for 
other vulnerabilities, e.g., reentrancy.
However, when incorrect access control is exploited, our approach is able to detect it based on its 
detrimental effects.

\begin{figure}[t]
	\centering
	\includegraphics[width=.85\columnwidth]{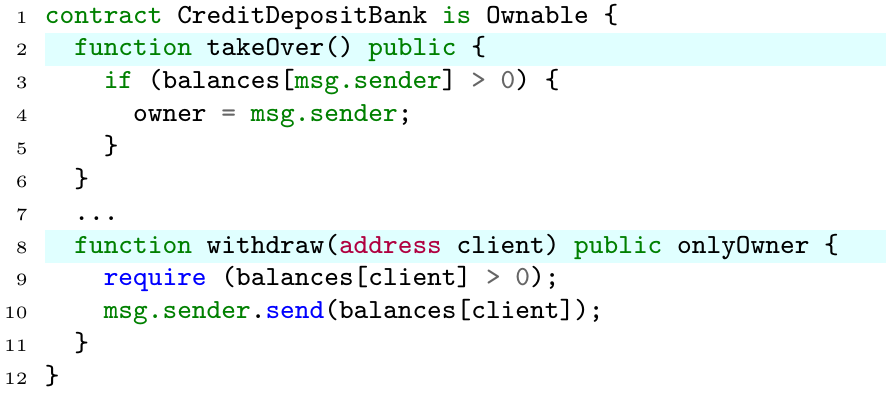}
	\caption{A vulnerability due to incorrect access control.}
	\label{fig:accessC}
\end{figure}

A code snippet from the smart contract \texttt{CreditDepositBank} (in Fig.~\ref{fig:accessC}) is vulnerable and was detected by our approach. 
The contract \texttt{CreditDepositBank} has a public function \texttt{takeOver()}, which can transfer ownership to the sender of the message.
A sender who acquires the ownership can then call the \texttt{withdraw()} function to withdraw \ether of any other participant.
Furthermore, the bookkeeping variable ``\texttt{balances}'' is not updated after the ``\texttt{send}'' (Line 10), thus violating the balance invariant.

\begin{figure}[t]
	\centering
	\includegraphics[width=.85\columnwidth]{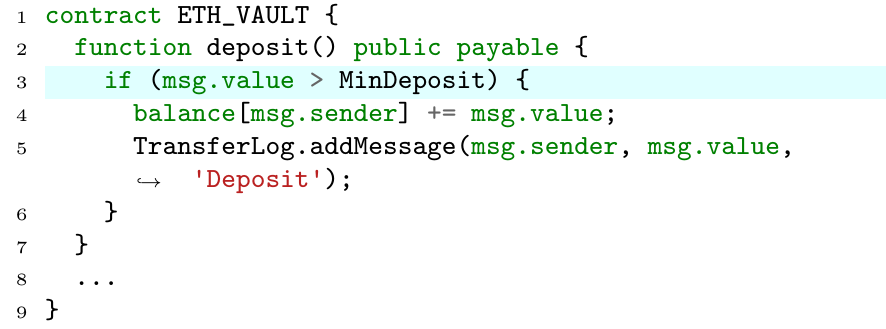}
	%
	\caption{A honey trap example.}
	\label{fig:honeyT}
\end{figure}

\paragraph{Honey Trap}
Some contracts, e.g, \texttt{ETH\_VAULT} and \texttt{WhaleGiveaway}, contain honey traps where the 
participants deposit \ether into the contracts, 
and cannot withdraw it again.
These smart contracts are unfair to the participants.
We take the code snippet from \texttt{ETH\_VAULT} in Fig.~\ref{fig:honeyT} to illustrate its mechanism.

The contract has a minimum deposit value, which is set by the owner of contract. 
When the participant deposits less \ether than the minimum value by invoking the \texttt{payable} function ``\texttt{deposit}'' 
(false branch at Line 3 in Fig.~\ref{fig:honeyT}), the \ether is still deposited into the contract without any record.
Therefore, the participant cannot withdraw his contribution again,
The transaction reduces the balance of participant without changing the bookkeeping variable.
This violates the transaction invariant.
The expected behavior should be to return the fund to the participant and roll back the transaction, if the contribution is less than the 
minimum requirement. 

\paragraph{Deposit Less and Withdraw More}
Some smart contracts, e.g., \texttt{BountyHunt}, \texttt{LZLCoin} and \texttt{PowerCoin}, are 
vulnerable by allowing an adversary to withdraw more \ether than they have deposited.

\begin{figure}[t]
	\centering
	\includegraphics[width=.85\columnwidth]{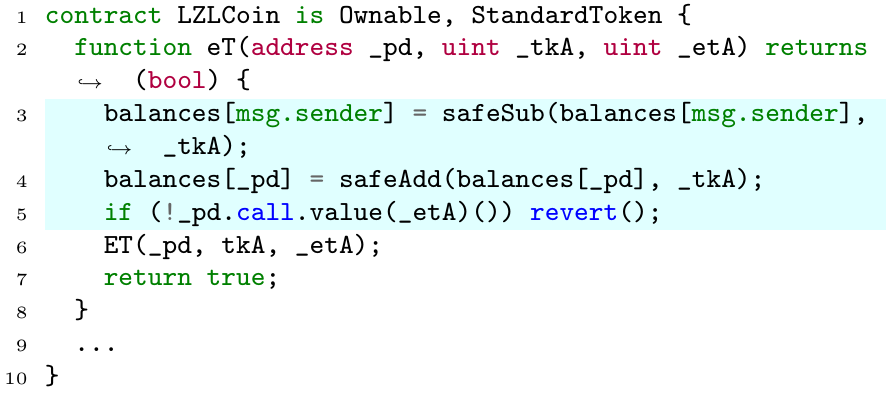}
	\caption{An example for deposit less and withdraw more.}
	\label{fig:ldmw}
\end{figure}

A code snippet from \texttt{LZLCoin} (Fig.~\ref{fig:ldmw}) illustrates this vulnerability.
The vulnerable function takes two parameters \texttt{\_tkA} and \texttt{\_etA}, which represent the 
balance a participant is able to withdraw and the \ether actually being sent out, respectively.
The values of these two parameters should always be equal, otherwise, the adversary is able to 
choose a smaller value for \texttt{\_tkA} and a larger value for \texttt{\_etA}, to withdraw more than pledged.
This behavior violates our balance invariant.

\subsubsection{Non-Exploitable Vulnerabilities Reported by the Pattern-Based Approaches}
\label{sec:non-exploitable}
In this section, we illustrate the reasons why some vulnerabilities detected by existing techniques are not exploitable.

\begin{figure}[t]
	\centering
	\includegraphics[width=.85\columnwidth]{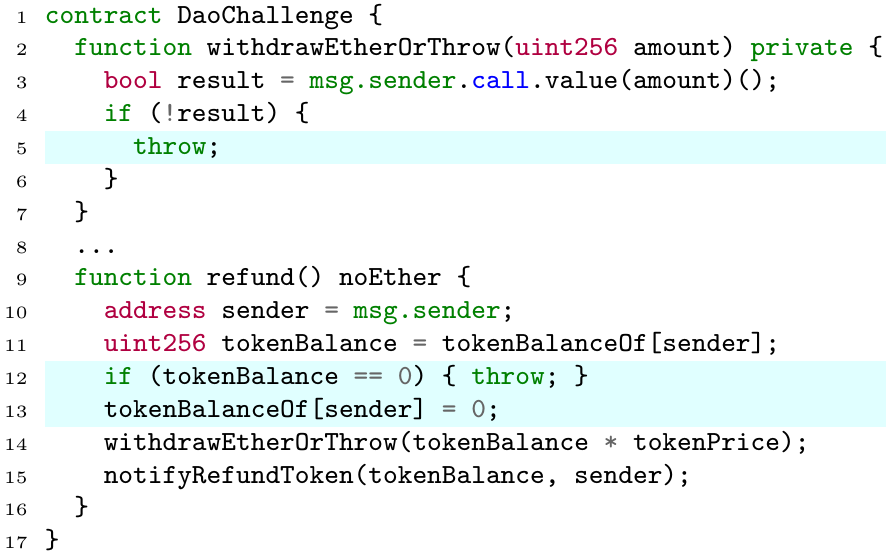}
	\caption{An example for non-exploitable reentrancy.}
	\label{fig:reentrancyE}
\end{figure}

\paragraph{Reentrancy}
Reentrancy describes the situation where a function is re-entered while in the midst of its execution. 
Reentrancy has become notorious due to the DAO bug~\cite{chang2018scompile}, and is considered as the top vulnerability~\cite{dasp}.
Based on predefined properties, existing techniques, e.\,g., Zeus, can detect reentrancy.
However, these properties, based on fixed patterns, are so strong that non-exploitable renentrancy is also reported.
For example, the contract \texttt{DaoChallenge} shown in Fig.~\ref{fig:reentrancyE} was reported by Zeus as vulnerable~\cite{kalra2018zeus}. 
However, based on the official website~\cite{daochallenge} where the contract was originally from, this is not an exploitable reentrancy.

This contract first checks whether the balance of the message sender is zero at Line $11$.
If so, it throws an exception and reverts the program state.
Otherwise, it sets the balance of the sender to zero at Line $12$, and then uses \texttt{withdrawEtherOrThrow()}, the safe withdraw function, to fetch 
\ether at Line $13$.
Through this safe withdraw function, the program may re-enter the function \texttt{refund()}.
When reentering function \texttt{refund()}, the balance will be set to zero.
Thus, the reentrancy cannot pass the check at Line $11$ again, and the program state is reverted.
As a result, an adversary cannot steal \ether from this contract.

\begin{figure}[t]
	\centering
	\includegraphics[width=.85\columnwidth]{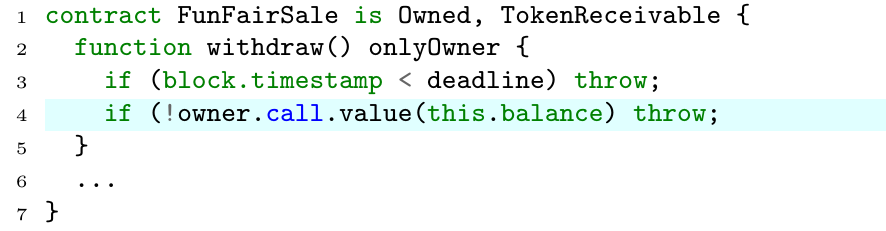}
	\caption{Another example for non-exploitable reentrancy.}
	\label{fig:reentrancyE2}
\end{figure}

Although ContractFuzzer adds extra conditions (e.g., a \texttt{call()} invocation has a greater-than-zero value) 
to reduce false positives, it still reports non-exploitable reentrancy cases. 
This is because the extra conditions are only based on syntax without considering the semantics.
For example, the code snippet from \texttt{FunFairSale} (Fig.~\ref{fig:reentrancyE2}) is a non-exploitable reentrancy detected by ContractFuzzer.
At Line $4$, the owner of the contract may withdraw all the balance.
Even if the \texttt{withdraw()} function is reentered, its program state is reverted due to not having enough funds to withdraw.

\paragraph{Exception Disorder}
An exception disorder occurs where there are inconsistencies in exception handling.
These inconsistencies are very hard to detect in smart contracts. 
ContractFuzzer uses a pattern to detect exception disorders: it checks if a \texttt{throw} statement is executed after a 
failed \texttt{send()}, in order to revert the transaction~\cite{jiang2018contractfuzzer}.
Zeus~\cite{kalra2018zeus} checks whether there is a write operation on a global variable after a failed \texttt{send()}. 
However, both checks are purely syntactic, and many non-exploitable vulnerabilities are reported because of this.

\begin{figure}[t]
	\centering
	\includegraphics[width=.85\columnwidth]{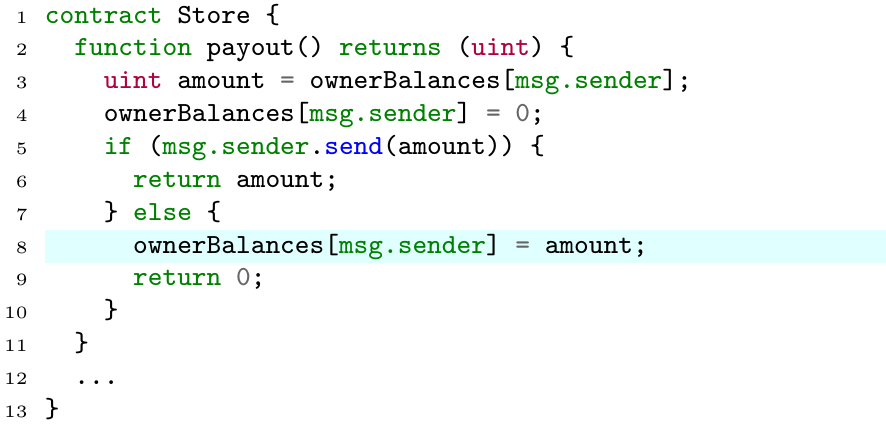}
	\caption{Non-exploitable exception disorder.}
	\label{fig:exceptionD}
\end{figure}

For example, consider the code snippet from \texttt{Store} in Fig.~\ref{fig:exceptionD}. 
At Line $5$, the contract pays out \ether to the message sender and the \texttt{send()} operation may fail.
When the \texttt{send()} operation fails, the contract reverts the program states at Line~$8$.
This is in fact a correct way to handle the exception. 
However, it is reported as a vulnerability by both ContractFuzzer and Zeus.
There is no easy way to precisely detect exception disorder without semantic understandings.

\paragraph{Integer Overflow/Underflow}
Integer overflow/underflow is an important issue in smart contracts, as they may allow adversaries 
to steal \ether or tokens from smart contracts.
The 1,095 out of 1,523 smart contracts in the dataset used by Zeus were reported as susceptible to integer overflow/underflow~\cite{integeroverflow}.
Although many integer overflow/underflows may occur in theory, not all of them are practical.
First, the total amount of \ether available on the Ethereum platform is limited to 140 million \ether~\cite{ethercap}.
Therefore, one cannot use infinite amount of \ether to overflow/underflow an \texttt{uint256} variable.
Second, the numbers of transactions and participants are also much smaller than the upper bound of \texttt{uint256}.
Finally, many smart contracts use safe mathematics operations protected from overfolw/underflow.
This is not recognized by the property(pattern)-based techniques, such as Zeus.

\begin{figure}[t]
	\centering
	\includegraphics[width=.85\columnwidth]{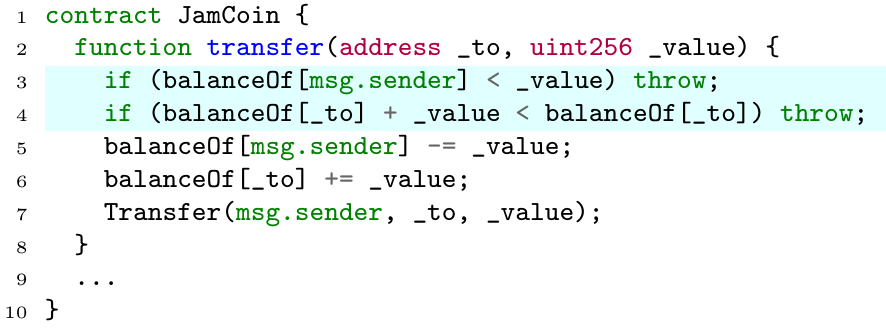}
	\caption{Non-exploitable integer overflow/underflow.}
	\label{fig:integerO}
\end{figure}

We take the code snippet from \texttt{JamCoin} (Fig.~\ref{fig:integerO}) as an example.
In this contract, before the integer operations at Lines $5-6$, it checks whether the operations would produce overflow/underflow at Lines $3-4$.
However, these checks are ignored by Zeus~\cite{kalra2018zeus}.
It requires program contexts being considered to accurately identify integer overflow/underflow.

\subsection{Threats to validity}
We have selected a benchmark set that is large enough to show the capabilities of \tool and compare it with other tools. However, both the set of benchmarks we selected ours from, as well as our own selection, may include a certain sample bias.
Thus, the results may not generalize to all smart contracts.
Moreover, while our approach is generic and also applicable to other types of smart contracts, 
some implementation details and issues found are specific to Ethereum.


%% file: relate.tex
\section{Related Work}
\label{sec:relate}
%
Existing work on smart contract vulnerability detection can be categorized into static 
analysis~\cite{tsankov2018securify,chang2018scompile,luu2016making,nikolic2018finding,kalra2018zeus,bhargavan2016formal}
 and dynamic analysis~\cite{jiang2018contractfuzzer,liu2018reguard}.

\subsection{Static Analysis}

\paragraph{Program Analysis}
Securify~\cite{tsankov2018securify} first infers semantic information by analyzing control- and data-dependencies 
of the contract code. Then, it checks against both the predefined compliance and violation properties to detect vulnerabilities.
SmartCheck~\cite{smartCheck} is an automated static code analyzer for smart contracts.
It automatically checks smart contracts against a knowledge base for security vulnerabilities and bad practices.
Slither~\cite{slither} is a static analysis framework for Solidity, 
which contains a suite of vulnerability detectors and also provides an API for developing custom analyses.

\paragraph{Symbolic Execution}
Oyente~\cite{luu2016making} is the first tool to apply symbolic execution in finding potential
vulnerabilities in smart contracts. It formulates the vulnerabilities as intra-procedural 
properties, and uses symbolic execution to check against these properties.
TEETHER~\cite{krupp2018teether} focuses their analysis on the critical paths of a contract program.
Specifically, a path is critical if it includes an instruction whose arguments can be controlled by an attacker.
Once a critical path is found, TEETHER computes the path conditions and infers the corresponding attack sequences 
for triggering the vulnerability.
In addition, TEETHER also requires that the value transmitted in the final \texttt{CALL} instruction is greater than the sum of all values sent to the contract.
This is similar to our approach but imprecise, because it does not model the whole transaction.
MAIAN~\cite{nikolic2018finding} is designed to find three types of problematic contracts: the prodigal, greedy and suicidal contracts. 
It formulates these three types of problems as inter-procedural properties, and performs bounded inter-procedural 
symbolic execution to search for property violations.
EthRacer~\cite{kolluri2019exploiting} investigates a family of event-ordering bugs in smart 
contracts.
These bugs are intimately related to the dynamic ordering of contract events, i.e., function calls.
The technical challenge in detecting event-ordering bugs in smart contract is the inherent 
combinatorial blowup in the path and state space analysis, even for simple contracts.
The authors propose to use partial-order reduction techniques, using automatically extracted 
\emph{happens-before} relations along with several dynamic symbolic execution optimizations.

\paragraph{Formal Verification}
There are also attempts to formally verify smart contracts using either model checking or 
theorem-proving~\cite{clarke2018model,kalra2018zeus,hirai2017defining,kevm,fstarlang,bhargavan2016formal,why3,permenevverx,harz2018towards,liu2019survey}.
Zeus~\cite{kalra2018zeus} first translates Solidity source code into LLVM~\cite{llvm} intermediate 
language, and then performs the verification with the SeaHorn verification framework~\cite{seahorn}.
Hirai~\cite{hirai2017defining} defines a formal semantic model for EVM using the Lem language, and 
proves safety properties of contract programs compiled to Lem, with the interactive theorem prover 
Isabelle/HOL.
KEVM~\cite{kevm} is a semantic encoding of EVM bytecode in the K-framework based on the rewriting logic.
VerX~\cite{permenevverx} is an automated verifier for proving functional properties of smart 
contracts.
VerX addresses an important problem, as all real-world contracts must satisfy custom functional
specifications.
VerX combines three techniques, enabling it to automatically verify temporal properties of infinite 
state smart contracts: (1) reduction of temporal property verification to reachability checking, 
(2) a new symbolic execution engine for EVM that is precise and efficient for a practical
fragment of smart contracts, and (3) delayed predicate abstraction which uses symbolic execution 
during transactions and abstraction at transaction boundaries.
VERISOL~\cite{lahiri2018formal} studies the safety and security of smart contracts in the Azure 
Blockchain Workbench, an enterprise Blockchain-as-a-Service offering from Microsoft.
It formalizes the \emph{semantic conformance} of smart contracts against a state machine model with 
access-control policies, and develops a highly-automated formal verifier for Solidity that can 
produce proofs as well as discover counterexamples.

Static analysis approaches can be more efficient in terms of running time, but they often suffer 
from high false-positive rate.
The main difference between our approach and these techniques is that, our approach dynamically 
executes the contract code on the real EVM environment, and therefore the detected vulnerabilities 
are guaranteed to be exploitable.

\subsection{Dynamic Analysis}
Some dynamic analysis techniques are proposed to stress the vulnerabilities of smart 
contracts~\cite{zheng2018detailed,feng2019precise}.

\paragraph{Input generation}
ContractFuzzer~\cite{jiang2018contractfuzzer} is a fuzzing framework for detecting vulnerabilities 
of Ethereum smart contracts. It proposes seven specific patterns for seven types of 
vulnerabilities. 
Based on these patterns, it generates fuzzing inputs, instruments the EVM to collect the execution 
traces, and analyzes the traces to identify vulnerabilities.
ReGuard~\cite{liu2018reguard} developed a fuzzing-based analyzer to automatically detect reentrancy 
vulnerabilities.
Specially, it performs fuzz testing on smart contracts by iteratively generating random but diverse transactions.
Based on the runtime traces, ReGuard dynamically identifies the reentrancy vulnerabilities.
Echidna~\cite{echidna} takes a contract program as well as a set of invariants as input, and 
generates random inputs to trigger potential vulnerabilities.
The invariants used by Echidna have to be written within the contract itself, thus they are not 
expressive enough to encode our inter-contract invariants.
Wüstholz and Christakis~\cite{wustholz2018learning} present a technique that extends greybox 
fuzzing with a method for learning new inputs based on already explored smart contract executions.
The learned inputs can be used to guide exploration towards specific executions, for instance, ones 
that increase path coverage.

The main difference between our approach and other dynamic approaches is that
we provide general principles that drill down to the very root of vulnerabilities,
while other approaches use generic properties to detect specific vulnerabilities.
By definition, fixed collections of properties are limited and modeled at the syntactic level; 
thus, they usually suffer from both false negatives and false positives.
We believe that the absence of a general and precise test oracle is the main reason that there 
exist very few dynamic techniques for detecting vulnerabilities in smart contracts.

\paragraph{Test oracle}
The lack of a precise test oracle is often the main bottleneck in software test automation~\cite{barr2015oracle}.
An empirical analysis of model-derived test cases for Java programs shows that using a test oracle 
roughly doubles the defect detection rate~\cite{LiO2014AEA}.
Most activities to support the test oracle focus on providing better specification mechanisms, or on mining properties from the documentation or comments~\cite{barr2015oracle}.
An \emph{implicit test oracle} covers assumptions that have to hold globally for well-defined 
applications, e.g., no memory access to unallocated or uninitialized memory should ever happen. 
Sereum~\cite{rodler2018sereum} protects the deployed smart contracts from being exploited. 
It addresses this problem in the context of re-entrancy exploits and propose a novel smart contract security technology, which protects existing, deployed contracts against re-entrancy attacks in a backwards compatible way based on run-time monitoring and validation.
Sereum does not require any modification to or any semantic knowledge of existing contracts.
Our work is also within that domain, as we cover the implicit assumption that no funds are created or destroyed by transactions. We therefore provide a valuable contribution in a field where it is in general very difficult to find useful implicit assumptions~\cite{barr2015oracle}.

In general, specifications are provided by developers, either on a case-by-case based in code, or as more general rules that apply throughout the program. Specifications can be provided as executable code in the form of software design requirement, as preconditions, invariants, and postconditions~\cite{meyer1992applying}. These facilities have been made available in the Solidity language as of version 0.4.10~\cite{best-practices,security-considerations}, but are not widely used yet.
Compared to verification on traditional platforms, these features on Solidity have the drawback that their usage incurs side effects (in terms of the gas cost of computing the expression being evaluated); in general, side effects should be avoided in such expressions~\cite{fritzson1994using}. 

%% file: conclusion.tex
\section{Conclusion}
\label{sec:conclude}

We propose \tool, a grey-box fuzzing approach for finding exploitable vulnerabilities in smart 
contracts.
Different from previous works, the proposed test oracle captures the very roots of 
transaction-related vulnerabilities based on invariants (Definitions~\ref{def:BalanceInvariant} and~\ref{def:TransactionInvariant}), which are essential and not specific to any particular attack pattern.
We also use feedback computed from the efficient runtime monitoring on EVM to guide the mutation of transaction sequences for fuzzing.
We have demonstrated that \tool is effective in finding exploitable vulnerabilities and produces much fewer false positives than the state-of-the-art.
Furthermore, we find and confirm three new attacks.


%% file: main_arxiv.bbl
\begin{thebibliography}{10}
\providecommand{\url}[1]{#1}
\csname url@samestyle\endcsname
\providecommand{\newblock}{\relax}
\providecommand{\bibinfo}[2]{#2}
\providecommand{\BIBentrySTDinterwordspacing}{\spaceskip=0pt\relax}
\providecommand{\BIBentryALTinterwordstretchfactor}{4}
\providecommand{\BIBentryALTinterwordspacing}{\spaceskip=\fontdimen2\font plus
\BIBentryALTinterwordstretchfactor\fontdimen3\font minus
  \fontdimen4\font\relax}
\providecommand{\BIBforeignlanguage}[2]{{%
\expandafter\ifx\csname l@#1\endcsname\relax
\typeout{** WARNING: IEEEtran.bst: No hyphenation pattern has been}%
\typeout{** loaded for the language `#1'. Using the pattern for}%
\typeout{** the default language instead.}%
\else
\language=\csname l@#1\endcsname
\fi
#2}}
\providecommand{\BIBdecl}{\relax}
\BIBdecl

\bibitem{bitcoin}
``{Bitcoin Project},'' \url{https://bitcoin.org/}, 2019.

\bibitem{ethereum}
``{Ethereum Project},'' \url{https://www.ethereum.org/}, 2019.

\bibitem{peters2016understanding}
G.~W. Peters and E.~Panayi, ``{Understanding Modern Banking Ledgers Through
  Blockchain Technologies: Future of Transaction Processing and Smart Contracts
  on the Internet of Money},'' in \emph{Banking Beyond Banks and Money}.\hskip
  1em plus 0.5em minus 0.4em\relax Springer, 2016, pp. 239--278.

\bibitem{xu2017design}
R.~Xu, L.~Zhang, H.~Zhao, and Y.~Peng, ``{Design of Network Media's Digital
  Rights Management Scheme Based on Blockchain Technology},'' in
  \emph{International Symposium on Autonomous Decentralized System}.\hskip 1em
  plus 0.5em minus 0.4em\relax IEEE, 2017, pp. 128--133.

\bibitem{iansiti2017truth}
M.~Iansiti and K.~R. Lakhani, ``{The Truth about Blockchain},'' \emph{Harvard
  Business Review}, 2017.

\bibitem{chang2018scompile}
J.~Chang, B.~Gao, H.~Xiao, J.~Sun, and Z.~Yang, ``{sCompile: Critical Path
  Identification and Analysis for Smart Contracts},'' \emph{arXiv preprint
  arXiv:1808.00624}, 2018.

\bibitem{tsankov2018securify}
P.~Tsankov, A.~Dan, D.~D. Cohen, A.~Gervais, F.~Buenzli, and M.~Vechev,
  ``{Securify: Practical Security Analysis of Smart Contracts},'' in \emph{ACM
  Conference on Computer and Communications Security}, 2018, pp. 67--82.

\bibitem{jiang2018contractfuzzer}
B.~Jiang, Y.~Liu, and W.~Chan, ``{ContractFuzzer: Fuzzing Smart Contracts for
  Vulnerability Detection},'' in \emph{Proceedings of the 33rd ACM/IEEE
  International Conference on Automated Software Engineering}.\hskip 1em plus
  0.5em minus 0.4em\relax ACM, 2018, pp. 259--269.

\bibitem{luu2016making}
L.~Luu, D.-H. Chu, H.~Olickel, P.~Saxena, and A.~Hobor, ``{Making Smart
  Contracts Smarter},'' in \emph{ACM Conference on Computer and Communications
  Security}.\hskip 1em plus 0.5em minus 0.4em\relax ACM, 2016, pp. 254--269.

\bibitem{nikolic2018finding}
I.~Nikolic, A.~Kolluri, I.~Sergey, P.~Saxena, and A.~Hobor, ``{Finding the
  Greedy, Prodigal, and Suicidal Contracts at Scale},'' \emph{arXiv preprint
  arXiv:1802.06038}, 2018.

\bibitem{kalra2018zeus}
S.~Kalra, S.~Goel, M.~Dhawan, and S.~Sharma, ``{Zeus: Analyzing Safety of Smart
  Contracts},'' in \emph{The Network and Distributed System Security
  Symposium}, 2018.

\bibitem{liu2018reguard}
C.~Liu, H.~Liu, Z.~Cao, Z.~Chen, B.~Chen, and B.~Roscoe, ``{ReGuard}: finding
  reentrancy bugs in smart contracts,'' in \emph{ACM/IEEE International
  Conference on Software Engineering}.\hskip 1em plus 0.5em minus 0.4em\relax
  ACM, 2018, pp. 65--68.

\bibitem{daoChallenge_source}
``{DaoChallenge Source},''
  \url{https://etherscan.io/address/0x80f1f62b8b365c5326100d462d8570771b8d0e57},
  2019, accessed 2019.

\bibitem{Feng2019PreciseAS}
Y.~Feng, E.~Torlak, and R.~Bod{\'\i}k, ``{Precise Attack Synthesis for Smart
  Contracts},'' \emph{arXiv preprint arXiv:1902.06067}, vol. abs/1902.06067,
  2019.

\bibitem{zalewski2016american}
M.~Zalewski, ``{American Fuzzy Lop},'' \url{http://lcamtuf.coredump.cx/afl/},
  2016.

\bibitem{grech2018madmax}
N.~Grech, M.~Kong, A.~Jurisevic, L.~Brent, B.~Scholz, and Y.~Smaragdakis,
  ``{Madmax: Surviving Out-of-Gas Conditions in Ethereum Smart Contracts},''
  \emph{Proceedings of the ACM on Programming Languages}, p. 116, 2018.

\bibitem{jakobsson1999proofs}
M.~Jakobsson and A.~Juels, ``{Proofs of Work and Bread Pudding Protocols},'' in
  \emph{Secure Information Networks}.\hskip 1em plus 0.5em minus 0.4em\relax
  Springer, 1999, pp. 258--272.

\bibitem{king2012ppcoin}
S.~King and S.~Nadal, ``{PPCoin: Peer-to-Peer Crypto-Currency with
  Proof-of-Stake},''
  \url{https://bitcoin.peryaudo.org/vendor/peercoin-paper.pdf}.

\bibitem{wu2018cream}
S.~Wu, Y.~Chen, Q.~Wang, M.~Li, C.~Wang, and X.~Luo, ``{CReam: A Smart Contract
  Enabled Collusion-Resistant E-Auction},'' \emph{IEEE Transactions on
  Information Forensics and Security}, vol.~14, no.~7, pp. 1687--1701, 2018.

\bibitem{solidity}
``{Solidity},'' \url{https://solidity.readthedocs.io/en/v0.5.1/}, 2018.

\bibitem{jiao2018executable}
J.~Jiao, S.~Kan, S.-W. Lin, D.~Sanan, Y.~Liu, and J.~Sun, ``{Executable
  Operational Semantics of Solidity},'' \emph{arXiv preprint arXiv:1804.01295},
  2018.

\bibitem{krupp2018teether}
J.~Krupp and C.~Rossow, ``{teEther: Gnawing at Ethereum to Automatically
  Exploit Smart Contracts},'' in \emph{27th USENIX Security Symposium USENIX
  Security}.\hskip 1em plus 0.5em minus 0.4em\relax ACM, 2018, pp. 1317--1333.

\bibitem{ERC20}
``{ERC-20},''
  \url{https://theethereum.wiki/w/index.php/ERC20\_Token\_Standard/}, 2018.

\bibitem{ERC721}
``{ERC-721},'' \url{http://erc721.org/}, 2019.

\bibitem{chatterjee2018quantitative}
K.~Chatterjee, A.~K. Goharshady, and Y.~Velner, ``{Quantitative Analysis of
  Smart Contracts},'' in \emph{European Symposium on Programming}.\hskip 1em
  plus 0.5em minus 0.4em\relax Springer, 2017, pp. 494--509.

\bibitem{atzei2017survey}
N.~Atzei, M.~Bartoletti, and T.~Cimoli, ``{A Survey of Attacks on Ethereum
  Smart Contracts},'' in \emph{Principles of Security and Trust}.\hskip 1em
  plus 0.5em minus 0.4em\relax Berlin, Heidelberg: Springer, 2017, pp.
  164--186.

\bibitem{wang2019vultron}
H.~Wang, Y.~Li, S.-W. Lin, L.~Ma, and Y.~Liu, ``{Vultron: Catching Vulnerable
  Smart Contracts Once and for All},'' in \emph{Proceedings of the 41st
  International Conference on Software Engineering: New Ideas and Emerging
  Results}.\hskip 1em plus 0.5em minus 0.4em\relax IEEE Press, 2019, pp. 1--4.

\bibitem{underflow}
``{Decentralized Application Security Project (or DASP) Top 10 of 2018},''
  \url{http://www.dasp.co/}, 2018, accessed 2018.

\bibitem{clarke2018model}
E.~M. Clarke~Jr, O.~Grumberg, D.~Kroening, D.~Peled, and H.~Veith, \emph{{Model
  Checking}}.\hskip 1em plus 0.5em minus 0.4em\relax MIT press, 2018.

\bibitem{listeelix}
Y.~Li, B.~Chen, M.~Chandramohan, S.-W. Lin, Y.~Liu, and A.~Tiu, ``{Steelix:
  Program-State Based Binary Fuzzing},'' in \emph{Proceedings of the 2017 11th
  Joint Meeting on Foundations of Software Engineering}.\hskip 1em plus 0.5em
  minus 0.4em\relax ACM, 2017, pp. 627--637.

\bibitem{kolluri2019exploiting}
A.~Kolluri, I.~Nikolic, I.~Sergey, A.~Hobor, and P.~Saxena, ``{Exploiting the
  Laws of Order in Smart Contracts},'' in \emph{Proceedings of the 28th ACM
  SIGSOFT International Symposium on Software Testing and Analysis}.\hskip 1em
  plus 0.5em minus 0.4em\relax ACM, 2019, pp. 363--373.

\bibitem{rawat2017vuzzer}
S.~Rawat, V.~Jain, A.~Kumar, L.~Cojocar, C.~Giuffrida, and H.~Bos, ``{VUzzer:
  Application-aware Evolutionary Fuzzing},'' in \emph{The Network and
  Distributed System Security Symposium}, vol.~17, 2017, pp. 1--14.

\bibitem{wood2018ethereum}
G.~Wood, ``{Ethereum: A Secure Decentralised Generalised Transaction Ledger.
  {Ethereum} Project Yellow Paper (2018)},'' 2018.

\bibitem{cfg}
H.~Wang, T.~Liu, X.~Guan, C.~Shen, Q.~Zheng, and Z.~Yang, ``{Dependence Guided
  Symbolic Execution},'' \emph{IEEE Transactions on Software Engineering},
  vol.~43, no.~3, pp. 252--271, 2017.

\bibitem{data-dependency}
J.~Ferrante, K.~J. Ottenstein, and J.~D. Warren, ``{The Program Dependence
  Graph and Its Use in Optimization},'' \emph{ACM Transactions on Programming
  Languages and Systems}, vol.~9, no.~3, pp. 319--349, Jul. 1987.

\bibitem{integeroverflow}
{Yoichi Hirai}, ``{Integer Overflow/Underflow},''
  \url{https://github.com/ethereum/solidity/issues/796\#issuecomment-253578925},
  2016, accessed 2019.

\bibitem{klees2018evaluating}
G.~Klees, A.~Ruef, B.~Cooper, S.~Wei, and M.~Hicks, ``{Evaluating Fuzz
  Testing},'' in \emph{Proceedings of the 2018 ACM SIGSAC Conference on
  Computer and Communications Security}.\hskip 1em plus 0.5em minus 0.4em\relax
  ACM, 2018, pp. 2123--2138.

\bibitem{dasp}
{NCC Group}, ``{DASP: Decentralized Application Security Project},''
  \url{https://www.dasp.co/}, 2019.

\bibitem{daochallenge}
{Sjors Provoost}, ``{Dao Challenge},''
  \url{https://medium.com/@dao.challenge/challenge-3-how-i-almost-lost-100-1a11a9824ccb},
  2016, accessed 2019.

\bibitem{ethercap}
{Vitalik Buterin}, ``{Ether Total Supply},''
  \url{https://github.com/ethereum/EIPs/issues/960}, 2018, accessed 2019.

\bibitem{bhargavan2016formal}
K.~Bhargavan, A.~Delignat-Lavaud, and e.~a. Fournet, ``{Formal Verification of
  Smart Contracts: Short Paper},'' in \emph{Proceedings of the 2016 ACM
  Workshop on Programming Languages and Analysis for Security}.\hskip 1em plus
  0.5em minus 0.4em\relax ACM, 2016, pp. 91--96.

\bibitem{smartCheck}
``{smartCheck},'' \url{https://tool.smartdec.net/}, 2019.

\bibitem{slither}
{Trail of Bits}, ``{Slither},'' \url{https://github.com/trailofbits/slither},
  2019, accessed 2019.

\bibitem{hirai2017defining}
Y.~Hirai, ``{Defining the Ethereum Virtual Machine for Interactive Theorem
  Provers},'' in \emph{International Conference on Financial Cryptography and
  Data Security}.\hskip 1em plus 0.5em minus 0.4em\relax Springer, 2017, pp.
  520--535.

\bibitem{kevm}
E.~Hildenbrandt, M.~Saxena, N.~Rodrigues, X.~Zhu, P.~Daian, D.~Guth, B.~Moore,
  D.~Park, Y.~Zhang, A.~Stefanescu, and G.~Rosu, ``{KEVM: A Complete Formal
  Semantics of the Ethereum Virtual Machine},'' in \emph{2018 IEEE 31st
  Computer Security Foundations Symposium}.\hskip 1em plus 0.5em minus
  0.4em\relax IEEE, 2018, pp. 204--217.

\bibitem{fstarlang}
{FStarLang}, ``Fstartlang,'' \url{https://www.fstar-lang.org/}, 2019, accessed
  2019.

\bibitem{why3}
{Why3}, ``Why3,'' \url{http://why3.lri.fr/}, 2019, accessed 2019.

\bibitem{permenevverx}
A.~Permenev, D.~Dimitrov, P.~Tsankov, D.~Drachsler-Cohen, and M.~Vechev,
  ``Verx: Safety verification of smart contracts,'' 2020.

\bibitem{harz2018towards}
D.~Harz and W.~Knottenbelt, ``Towards safer smart contracts: A survey of
  languages and verification methods,'' \emph{arXiv preprint arXiv:1809.09805},
  2018.

\bibitem{liu2019survey}
J.~Liu and Z.~Liu, ``{A Survey on Security Verification of Blockchain Smart
  Contracts},'' \emph{IEEE Access}, 2019.

\bibitem{llvm}
{The LLVM Foundation}, ``{LLVM},'' \url{https://llvm.org/}, accessed 2019.

\bibitem{seahorn}
``{SeaHorn},'' \url{http://seahorn.github.io/}, 2019.

\bibitem{lahiri2018formal}
S.~K. Lahiri, S.~Chen, Y.~Wang, and I.~Dillig, ``{Formal Specification and
  Verification of Smart Contracts for Azure Blockchain},'' \emph{arXiv preprint
  arXiv:1812.08829}, 2018.

\bibitem{zheng2018detailed}
P.~Zheng, Z.~Zheng, X.~Luo, X.~Chen, and X.~Liu, ``{A Detailed and Real-Time
  Performance Monitoring Framework for Blockchain Systems},'' in \emph{2018
  IEEE/ACM 40th International Conference on Software Engineering: Software
  Engineering in Practice Track (ICSE-SEIP)}.\hskip 1em plus 0.5em minus
  0.4em\relax IEEE, 2018, pp. 134--143.

\bibitem{feng2019precise}
Y.~Feng, E.~Torlak, and R.~Bodik, ``{Precise Attack Synthesis for Smart
  Contracts},'' \emph{arXiv preprint arXiv:1902.06067}, 2019.

\bibitem{echidna}
{Trail of Bits}, ``{Echidna},'' \url{https://github.com/trailofbits/echidna},
  2019, accessed 2019.

\bibitem{wustholz2018learning}
V.~W{\"u}stholz and M.~Christakis, ``{Learning Inputs in Greybox Fuzzing},''
  \emph{arXiv preprint arXiv:1807.07875}, 2018.

\bibitem{barr2015oracle}
E.~T. Barr, M.~Harman, P.~McMinn, M.~Shahbaz, and S.~Yoo, ``{The Oracle Problem
  in Software Testing: A Survey},'' \emph{IEEE transactions on software
  engineering}, vol.~41, no.~5, pp. 507--525, 2015.

\bibitem{LiO2014AEA}
N.~Li and J.~Offutt, ``{An Empirical Analysis of Test Oracle Strategies for
  Model-Based Testing},'' in \emph{IEEE Seventh International Conference on
  Software Testing, Verification and Validation}, March 2014, pp. 363--372.

\bibitem{rodler2018sereum}
M.~Rodler, W.~Li, G.~O. Karame, and L.~Davi, ``Sereum: Protecting existing
  smart contracts against re-entrancy attacks,'' 2019.

\bibitem{meyer1992applying}
B.~Meyer, ``{Applying `Design by Contract'},'' \emph{Computer}, vol.~25, pp.
  40--51, 1992.

\bibitem{best-practices}
``{Smart Contract Security Best Practices},''
  \url{https://github.com/ConsenSys/smart-contract-best-practices}, 2019,
  accessed 2019.

\bibitem{security-considerations}
{Ethereum}, ``{Security Considerations},''
  \url{https://solidity.readthedocs.io/en/develop/security-considerations.html},
  2019, accessed 2019.

\bibitem{fritzson1994using}
P.~Fritzson, M.~Auguston, and N.~Shahmehri, ``{Using Assertions in Declarative
  and Operational Models for Automated Debugging},'' \emph{Journal of Systems
  and Software}, vol.~25, no.~3, pp. 223--239, 1994.

\end{thebibliography}
